\documentclass[12pt]{iopart}

\usepackage{graphicx}%
\usepackage{multirow}%
\makeatletter
\expandafter\let\csname equation*\endcsname\@undefined
\expandafter\let\csname endequation*\endcsname\@undefined
\makeatother
\usepackage{amsmath, amssymb, amsfonts}%
\usepackage{amsthm}%
\usepackage{siunitx}
\usepackage[title]{appendix}%
\usepackage{xcolor}%
\usepackage{textcomp}%
\usepackage{manyfoot}%
\usepackage{booktabs}%
\usepackage{algpseudocode}%
\usepackage{listings}%
\usepackage{caption}
\usepackage{subcaption}
\usepackage{colortbl}

\newcommand{\twocolumequationpair}[4]{%
  \begin{subequations}
  \renewcommand{\theequation}{\theparentequation\alph{equation}}
  \vspace{0.5em}
  \noindent
  \makebox[0.48\linewidth][c]{Speedmeter}
  \hfill
  \makebox[0.48\linewidth][c]{Position meter}

  \vspace{-2em}

  \noindent
  \begin{minipage}[t]{0.48\linewidth}
    \begin{align}
      #1 \label{#2}
    \end{align}
  \end{minipage}%
  \hfill
  \vrule width 0.5pt
  \hfill
  \begin{minipage}[t]{0.48\linewidth}
    \begin{align}
      #3 \label{#4}
    \end{align}
  \end{minipage}
  \vspace{1em}

  \end{subequations}
}

\usepackage{cancel}
\usepackage{enumitem}
\DeclareMathAlphabet{\mathcal}{OT1}{pzc}{m}{it}
\newlist{inlinelist}{itemize*}{1}
\setlist*[inlinelist,1]{label=\textbullet,
                        itemjoin={{ \ }}}

\renewcommand{\Re}{\mathop{\rm Re}\nolimits}

\begin{document}
\title{Simultaneous Speedmeter and Position-Meter Response in a Single Tabletop Interferometer}

\author{Mikhail Korobko}
\vspace*{-3mm}\ead{mikhail.korobko@uni-hamburg.de}
\address{Institute for Quantum Physics \& Center for Optical Quantum Technologies, University of Hamburg, Luruper Chaussee 149, 22761 Hamburg, Germany}\vspace*{-5mm}
\author{Xiang Li}\vspace*{-1mm}
\address{ASML, Silicon Valley - Building 2, 125 Rio Robles, San Jose, CA 95134, US}
\author{Torben Sobottke}\vspace*{-1mm}
\address{Institute for Quantum Physics \& Center for Optical Quantum Technologies, University of Hamburg, Luruper Chaussee 149, 22761 Hamburg, Germany}\vspace*{-5mm}
\author{Yiqiu Ma}\vspace*{-1mm}
\address{National Gravitation Laboratory, Hubei Key Laboratory of Gravitation and Quantum Physics,
School of Physics, Huazhong University of Science and Technology, Wuhan, 430074, China}
\address{Department of Astronomy, School of Physics, Huazhong University of Science and Technology, Wuhan, 430074, China}
\author{Yanbei Chen}\vspace*{-1mm}
\address{Burke Institute of Theoretical Physics, California Institute of Technology, Pasadena, CA, 91125, USA}
\author{Roman Schnabel}\vspace*{-1mm}
\address{Institute for Quantum Physics \& Center for Optical Quantum Technologies, University of Hamburg, Luruper Chaussee 149, 22761 Hamburg, Germany}
\vspace{10pt}
\begin{indented}
\item[]July 22, 2025
\end{indented}

\begin{abstract}

Quantum radiation-pressure noise (QRPN) limits the low-frequency sensitivity of gravitational wave detectors. 
The established method for suppressing QRPN is the injection of frequency-dependent squeezed light. It requires long-baseline filter cavities introducing substantial experimental complexity. A completely different interferometer concept is the speedmeter. It avoids QRPN at the source by measuring test mass speed instead of position. While extensively researched theoretically, speedmeters are yet to be demonstrated with a moving test mass in an optomechanical setting.
In this work, we present the first experimental observation of speedmeter behavior in a system with a movable test mass.
We realize a novel hybrid readout cavity configuration that enables simultaneous extraction of position and speed signals from two distinct output ports. We compare the optical transfer functions associated with each channel and observe the expected scaling behavior that distinguishes a speedmeter from a position-meter. We support our observations with a detailed theoretical model, showing how the hybrid readout cavity implements key speedmeter features.
Our results underscore the relevance of the speedmeter concept as an alternative for mitigating QRPN in future detectors and lay the groundwork for further experimental exploration.
\end{abstract}

%
%
%
%
%

\section{Introduction}
The sensitivity of current gravitational wave (GW) detectors is limited by quantum noise\,\cite{Caves1981}.  It arises from quantum uncertainties of the laser field used to measure the differential position of the test masses. The uncertainty in the field's phase quadrature produces photon shot noise (SN). The uncertainty in the field's amplitude quadrature produces quantum radiation-pressure noise (QRPN)\,\cite{Caves1980}. The established strategy to reduce both noise contributions relies on the injection of squeezed light\,\cite{Caves1981,Schnabel2010,LSC2011,Grote2013,Schnabel2017}, with an optimized, frequency-dependent squeezing angle\,\cite{Unruh1983,Jaekel1990} implemented using long-baseline, low-linewidth filter cavities\,\cite{Kimble2001}. All recent observations of GWs were performed on squeezed-light enhanced detectors\,\cite{GWo3a2021}. While effective, this approach becomes increasingly complex and costly as sensitivity demands rise—particularly for future detectors like Cosmic Explorer\,\cite{evansCosmicExplorerSubmission2023, hallCosmicExplorerNextGeneration2022} and Einstein Telescope\,\cite{abacScienceEinsteinTelescope2025, Abbott2016a, branchesiScienceEinsteinTelescope2023, Maggiore2019, korobkoQuantumTechnologiesEinstein2025}.

An alternative to this approach is the `speedmeter'. It is an interferometer configuration that senses differential test mass speed rather than differential position, as performed in Michelson and Mach-Zehnder interferometers.
Changing the measurement process allows to naturally mitigate QRPN, which originates in the measurement back-action through the light's radiation pressure.
In a speedmeter, the speed measurement is realized by two successive measurements of position with opposite sign, associated with two opposite radiation pressure uncertainty contributions.
This leads to suppression of QRPN without squeezed light injection.
The additional squeezed light injection suppresses quantum noise in a broad band without requiring a frequency dependent squeeze angle, since in a speedmeter both SN and QRPN share the same frequency scaling. This sets it apart from position meters and makes it a compelling concept for future GW detector designs\,\cite{Danilishin2012, Danilishin2019}.

While conceptually simple, realising a speedmeter for GW detection is non-trivial. 
Multiple speedmeter topologies have been proposed and theoretically investigated since the conception of this topology in the early 90s\,\cite{Braginsky1990, 96a1KhLe}, including designs based on sloshing cavities\,\cite{Braginsky2000a, 02a2Kh, Chen2003, Purdue2001, Huttner2017a}, Sagnac interferometers\,\cite{04a1Da, purduePracticalSpeedMeter2002, Barr, Huttner2017a, zhangQuantumNoiseCancellation2018, spencerExperimentalInvestigationLimitations2021}, dual-polarization schemes\,\cite{barinovIncreasingQuantumSpeed2025, Danilishin2018}, and EPR entanglement techniques\,\cite{knyazevSpeedmeterSchemeGravitationalwave2018}, among others\,\cite{Korobko2015, vyatchaninQuantumSpeedMeter2016, nishinoTeleportationbasedSpeedMeter2025}.
However, out of all these approaches only one experimental demonstration was presented, and that without a movable test mass\,\cite{DeVine2003a}. 
Therefore, the speedmeter has remained largely a theoretical concept.

In this work, we present the first direct experimental demonstration and comparison of the optical and optomechanical responses of a speedmeter and a position meter containing a movable mass.
This was made possible by introducing a novel speedmeter configuration in which semi-transparent the test mass is placed inside a ring cavity, forming a hybrid readout cavity (HRC). This setup directly implements the core principle of the speedmeter -- two sequential position measurements with a phase offset to extract speed -- and uses the cavity to enhance the resulting signal. This configuration also introduces a distinctive form of optomechanical coupling, which has been termed \textit{coherent}\,\cite{khaliliGeneralizedAnalysisQuantum2016, Li2019}. We show theoretically that our configuration yields the expected speedmeter behavior and discuss how it could be adapted for use in GW detectors. 
Our work is the first experimental observation of speedmeter characteristics in a real optomechanical system, and the comparison to the position meter highlights the promise of a speedmeter concept as an alternative to standard quantum noise suppression techniques.

\section{Experimental comparison of speed and position measurement}
\subsection{Expected signatures of a speedmeter}
The key signature of a speedmeter is the scaling behavior of its optical transfer function, which underpins all quantum properties that make it a promising candidate for future gravitational wave detectors. 
The optical transfer function defines how strongly the optical quantum field couples to the test masses.
It describes how much information about the test mass motion is transferred to the light beam and vice versa, which can be described for each Fourier (signal) frequency $\Omega$ by the optomechanical coupling factor $\mathcal{K}(\Omega)$\cite{Kimble2001}. It thus defines the strength of the optical signal as well as back action noise.
If quantum correlations are not exploited, it defines by how much the sensitivity of a GW detector exceeds the standard quantum limit (SQL), expressed in terms of one-sided strain-normalized power spectral density:
\begin{equation}\label{eq:general_sens}
S_h(\Omega) = \frac{S_{\rm SQL}(\Omega)}{2}\left(\frac{1}{\mathcal{K}(\Omega)} + \mathcal{K}(\Omega)\right)\, ,
\end{equation}
where $S_{\rm SQL} = 8\hbar / (L^{2}M^{1}\Omega^{2})$, with $L$ the interferometer arm length, $M$ is the reduced mass of the test masses, and $\Omega$ is the angular sideband frequency, i.e.~the angular signal frequency.
The SQL defines the lowest achievable signal-normalized noise for a detector that doesn't employ quantum correlations. 
$1/\mathcal{K}(\Omega)$ and $\mathcal{K}(\Omega)$ represent the contributions from SN and QRPN, respectively.

The coupling factor $\mathcal{K}(\Omega)$ has different characteristics for speedmeter and position meter.
For signal frequencies well below the optical linewidth of any cavities in the setup, the coupling factor behaves approximately as:\\[2mm]

\twocolumequationpair
{\mathcal{K}(\Omega)\sim \frac{P\tau^2}{M},}{eq:velocity:scale}
{\mathcal{K}(\Omega)\sim \frac{P}{M\Omega^2},}{eq:position:scale}
\noindent where $\tau$ is the characteristic time delay 
between the two sign-flipped position measurements in the speedmeter and $P$ is the total power at the test masses.
The key difference is evident: in a speedmeter, $\mathcal{K}(\Omega)$ is independent of frequency at low $\Omega$, meaning SN and QRPN scale identically with frequency. In contrast, for a position meter, $\mathcal{K}(\Omega)$ is frequency-dependent, resulting in different 
frequency-dependencies of SN and QRPN, which requires frequency-dependent squeezed light to achieve broadband quantum noise reduction.

\begin{figure}
  \centering
  \includegraphics[width=0.6\textwidth]{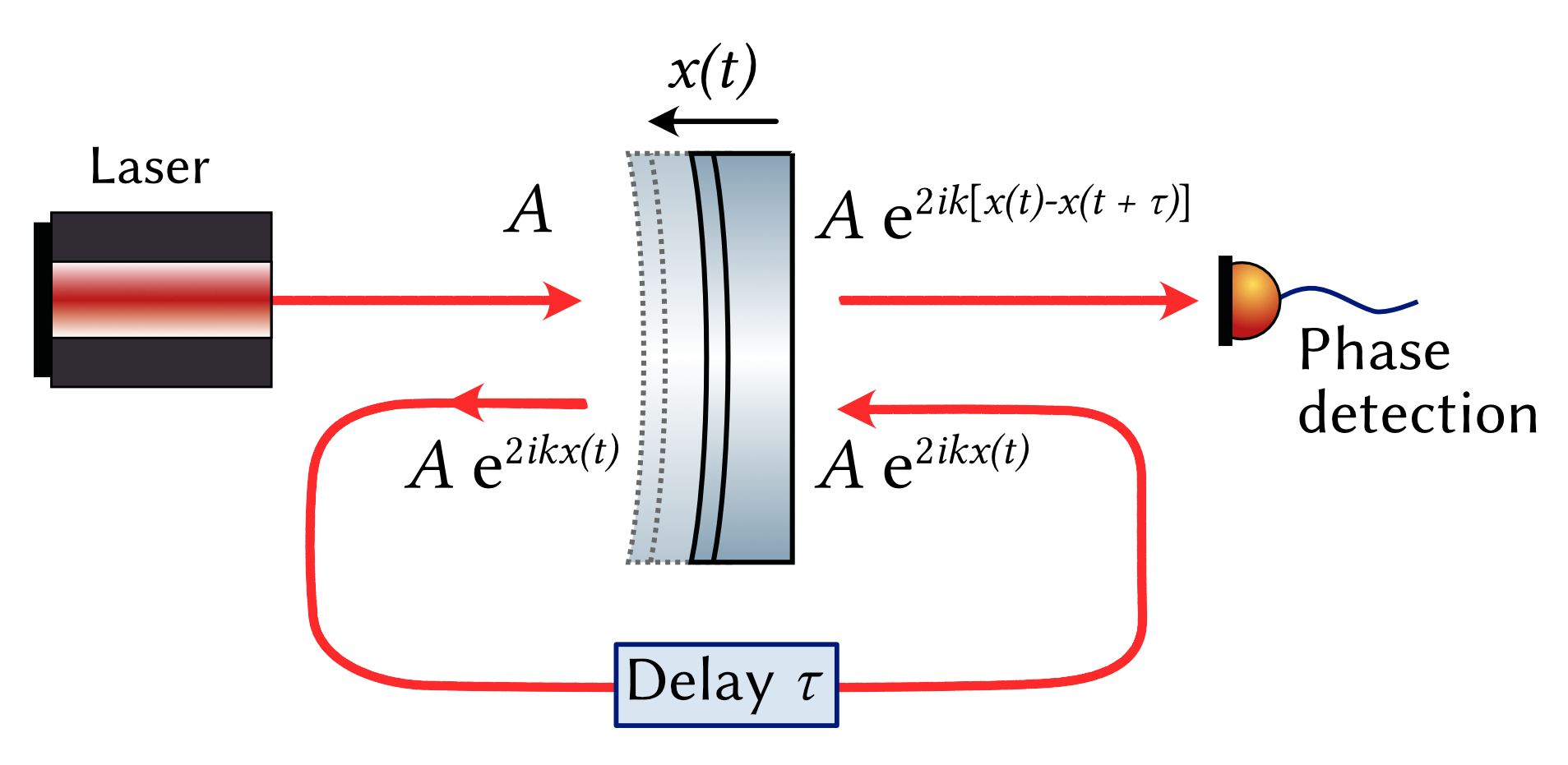}
  \caption[The speedmeter concept.]
  {The speedmeter concept\,\cite{Danilishin2012}. 
  The motion of the mirror $x(t)$ is sensed by light field with amplitude $A$ from two sides with time delay $\tau$.  On the first reflection it acquires a phase $2ikx(t)$, and on the second a phase $2ikx(t+\tau)$, where $k$ is the wavenumber.
  The signal on the phase-sensitive detector (e.g.\,a homodyne detector) is then proportional to $x(t)-x(t+\tau) = v(t+\tau)\tau$. Since the light field's momentum transfer to the mirror is the same from both sides.
  The equality includes the quantum uncertainty, and the net radiation-pressure force approaches zero for $\tau$ approaching zero. A speedmeter is always a `quantum speedmeter'.} 
  \label{fig:1}
\end{figure}

The simplest way to realize a speedmeter is to reflect 
the same light beam off the same movable mirror twice. 
If this is done from opposite sides, QNRP is strongly mitigated (see Fig.\,\ref{fig:1}). A `speed signal' is generated through the repeated measurement. 
The first reflection produces a phase shift that depends on the mirror's position $x(t)$. 
On the second reflection, occurring after a delay $\tau$, the light acquires a phase shift of $-x(t+\tau)$. The total phase shift is then proportional to the mirror's speed $v(t)$:
\begin{equation}
  x(t) - x(t+\tau) \approx v(t+\tau)\,\tau \, .
\end{equation}
The QRPN imprinted on the test mass is proportional to $F_{\rm QRPN, sm}\sim a(t)-a(t+\tau)$, where $a(t)$ is the fluctuating quantum amplitude of the light field.
Momentum transfer from this field to one side of the mirror partially compensates the moment transfer to the other side, and the resulting QRPN is suppressed.
In frequency domain, $F_{\rm QRPN, sm}(\Omega)\sim \Omega \tau a(\Omega) = F_{\rm QRPN, pm}\Omega \tau$, which is scaled by $\Omega \tau \ll 1$ compared to a standard position meter QRPN $F_{\rm QRPN, pm}$.
The simple argument above describes the speedmeter's key feature: Reading out the speed mitigates QRPN, which is the motivation for using the speedmeter concept in GW detection.
The quantitative derivation of the speedmeter behavior is presented in a later section.

\subsection{Optomechanical hybrid readout cavity as a speedmeter}
\begin{figure}
  \includegraphics[width=\textwidth]{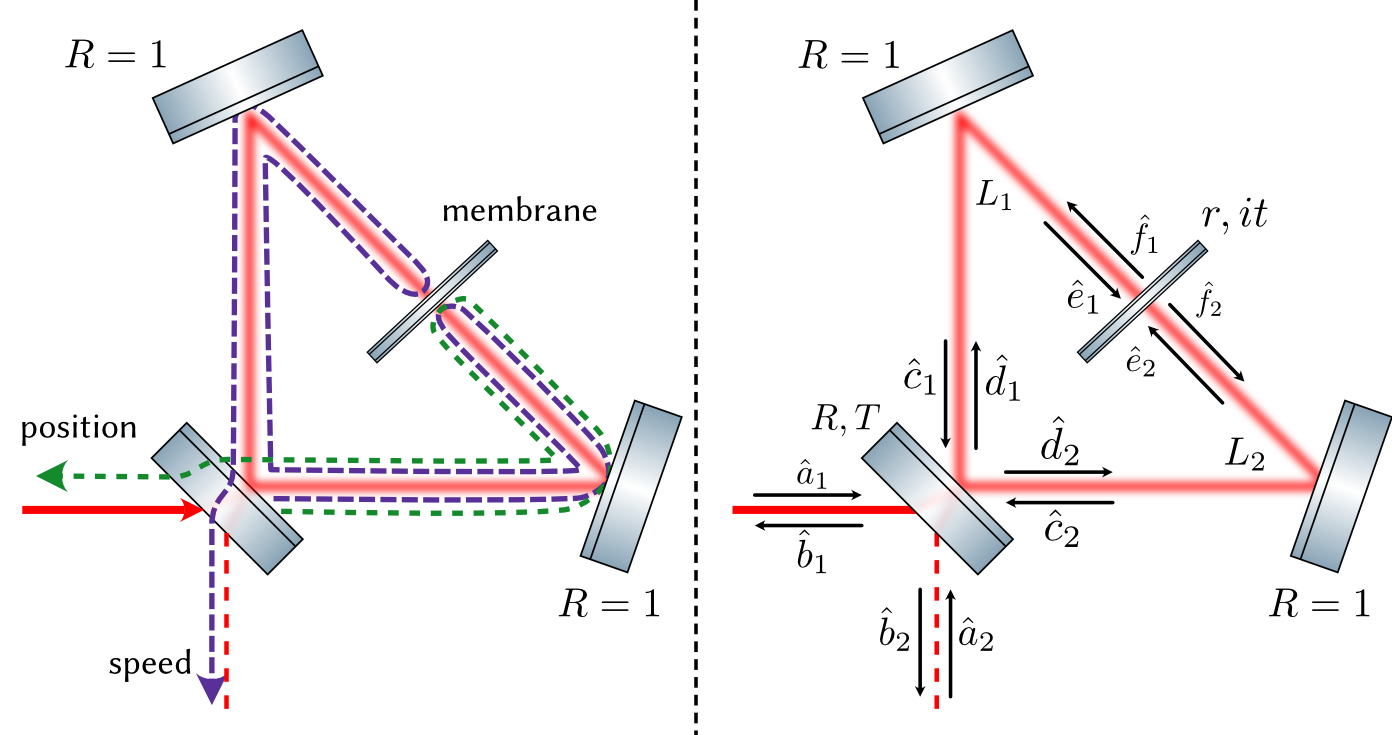}
  \caption[Optomechanical hybrid readout cavity.]
  {
Optomechanical hybrid readout cavity (HRC) --- 
Left: Light is mode-matched to the HRC through a semi-transparent coupling mirror. The same mirror establishes two output ports. Our analysis shows that all the carrier light is back-reflected towards the laser source together with information about the membrane {\it position}. The other port does not contain any carrier light, see Eq.\,(\ref{eq:17}) and solely information about the membrane {\it speed}.
Right: Notation, including the complex amplitudes of the light field.
}
  \label{fig:2}
\end{figure}

Our optomechanical hybrid readout cavity (HRC) is a rigid three-mirror cavity containing a movable semi-transparent SiN membrane\,\cite{Jayich2008a} serving as a test mass, see Fig.\,\ref{fig:2}. The light is coupled to the cavity via one of the rigidly mounted cavity mirrors.
A ring cavity naturally supports two types of modes, clock-wise and counter-clock-wise traveling modes. In our optomechanical HRC only one mode is excited by laser light, but the membrane reflection excites also the counter-propagating one.
As we show below, this results in a particular resonance structure and a dark output port that contains only speed information with neither position information nor carrier light.
We also note that the motion of the membrane redistributes the energy between the two modes, which leads to an unusual type of optomechanical coupling, called ``coherent''\,\cite{khaliliGeneralizedAnalysisQuantum2016} and studied in detail in application to the HRC in Ref.\,\cite{Li2019}.

Our HRC in Fig.\ref{fig:2} is a realistic extension of a thought experiment in Fig.\ref{fig:1}, where the light path after two opposing reflections off the test mass is looped onto itself. 
The symmetry of the setup allows to separate the signal from the test mass into two output ports, which we refer to as \emph{speed} and \emph{position} outputs.
Upon entering the cavity, the beam is either reflected or transmitted through the test mass.
After one cavity roundtrip, the beam is again reflected or transmitted through the membrane test mass.
The beam that is reflected twice acquires speed information --- in a direct analogy with Fig.\,\ref{fig:1}.
This beam exits through the speed output port, together with contributions from all even number of reflections.
The port contains purely speed information.
Position information as well as carrier light interferes destructively.
The beam that is reflected an odd amount of times is retro-reflected towards the laser source exiting the position output port.
This beam contains a combination of the position information and speed information.

\subsection{Experimental comparison of speed and position responses}
\begin{figure}[htb!]
  \includegraphics[width=1\textwidth]{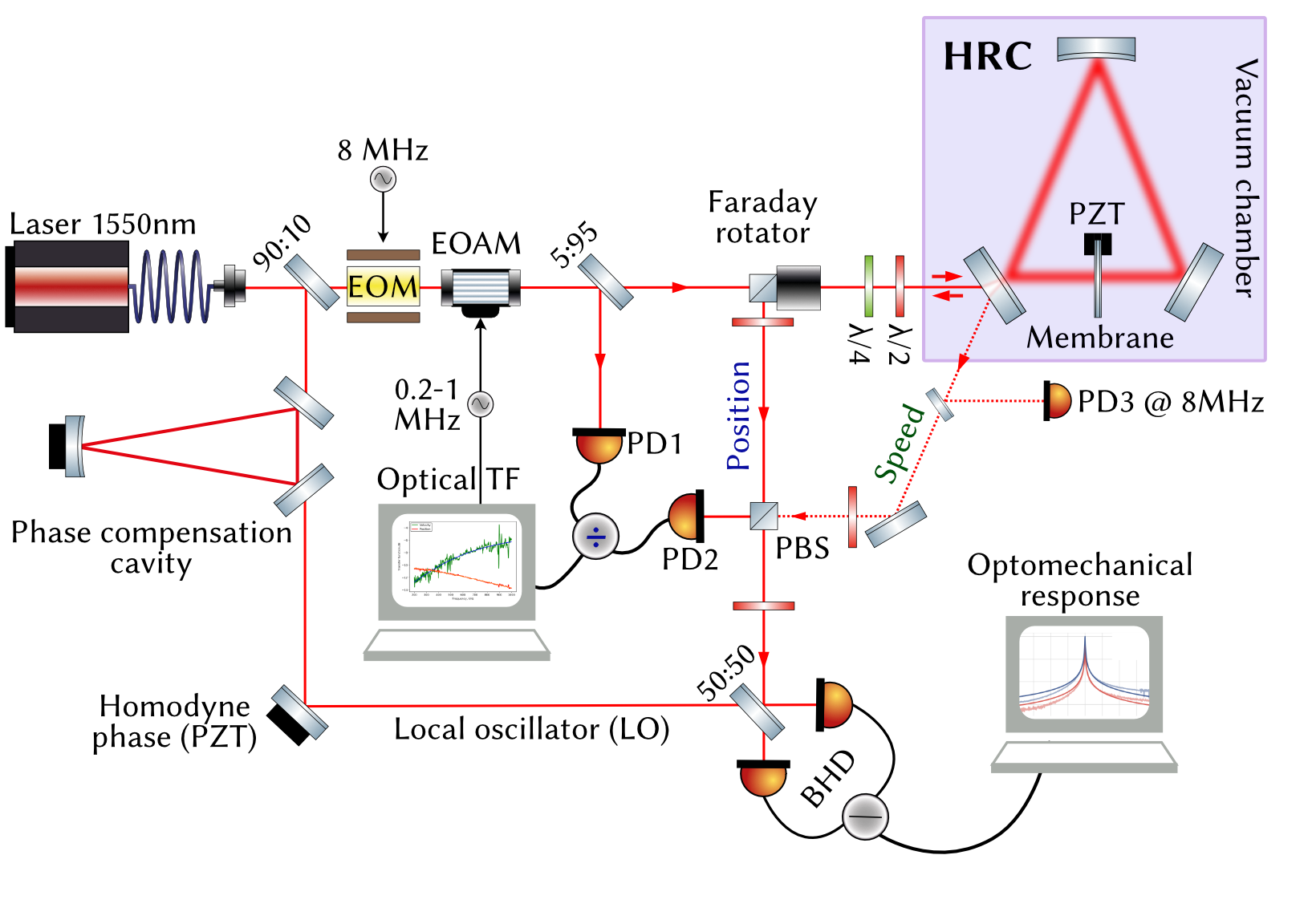}
  \caption[Simplified setup of our experiment]{Simplified setup of our experiment.
The output of a 1550-nm fibre laser was split into a local oscillator (LO) and a sensing beam. 
  The LO passed a phase compensation cavity, which also reduced the phase noise and cleaned the LO's spatial mode profile. 
  The sensing beam passed an electro-optic amplitude modulator (EOAM) for probing the optical transfer function of the HRC inside the vacuum chamber. It produced amplitude modulation sidebands at frequencies 0.2-1\,MHz that were detected on PD1 and PD2. The electro-optic modulator (EOM) in the same path generated phase modulation sidebands that were used for the stabilization of the laser frequency to a resonance of the HRC, to which the sensing beam was mode-matched. PD3 was used for stabilization control.
  The retro-reflected measurement beam (from the `position port') passed a Faraday isolator and a polarising beam splitter (PBS) for separating it from the incoming beam.
Both reflected measurement beams (including that from the `speed port')
  were overlapped with the LO beam on a balanced beam splitter for balanced homodyne detection (BHD). 
BHD recorded the signal from the membrane excited by an external force (PZT), thus measuring the optomechanical response of the system. 
  The phase compensation cavity was used to match the optical path length for sensing beam and LO and thus partially cancel laser frequency noise.
$\lambda/2$: Half wave plate. $\lambda/4$: Quarter wave plate. PD: Photo diode. TF: Transfer function. PZT: Piezo electric element.
  }
  \label{fig:3}
\end{figure}

Fig.\,\ref{fig:3} shows the schematic of our experiment. Its innovative component was the optomechanical HRC in the top right corner. It contained a micro-mechanical SiN membrane serving as a movable mirror. Its reflectivity at 1550\,nm was  $|r|^2 \approx 4.6\%$.
The reflectivity coupled the two counter propagating modes, which resulted in normal-mode splitting (see subsection \ref{ssec:resonances}). 
The cavity linewidth was $\gamma/2\pi = 0.84\pm 0.01$\,MHz and the fundamental resonance frequency of the membrane $\approx 0.4$\,MHz.
When the continuous-wave 1550-nm beam was coupled to one of the cavity resonances, the mechanical membrane resonance produced a strong peak in the spectrum of both cavity output beams when measured with the BHD.
For this,
the two output fields were recombined on a polarizing beamsplitter, allowing to propagate them along the same path and measure on the same BHD. The specific output then was selected by blocking either of the ports and adjusting the polarization to match the operating point of the homodyne detector.
Both the optical cavity length and the relative phases between the measurement beams and the local oscillator in the homodyne detector were actively stabilized.

\begin{figure}
  \centering
  \includegraphics[width=0.9\textwidth]{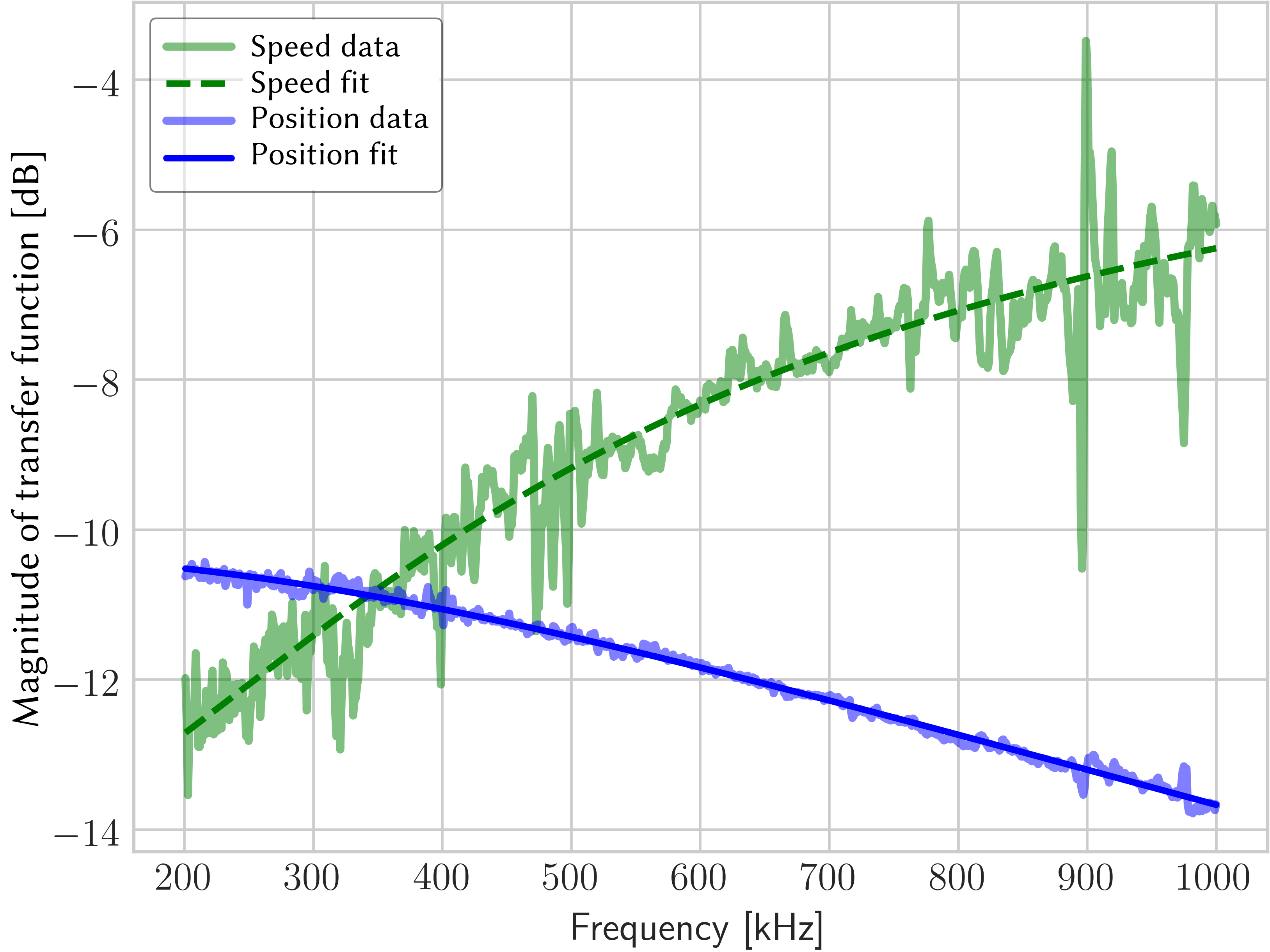}
  \caption[Measurement of the optical transfer function of the HRC on the position and speed ports.]
  {
  Measurement of the optical transfer function of the HRC on the position and speed ports. The transfer function was measured by observing the change in the amplitude modulation sidebands imprinted on the light fields as they go through the HRC. The frequency scaling of the position and speed port is clearly different and matches well with the theoretical expectations. The fit of the theory (see the Appendix) gives an optical linewidth of $0.90 \pm 0.05$\,MHz, which matches the independently measured value of approximately $0.84\pm 0.01$\,MHz.
}
  \label{fig:4}
\end{figure}

We observed two speedmeter signatures from the HRC. 
First, we demonstrated the speedmeter optical response by measuring transfer functions of optical amplitude modulations from the input port to the two HRC output ports. The amplitude modulation signal was produced with an electro-optical amplitude modulator (EOAM) and measured on the two output ports by two photo diodes. By sweeping the modulation frequency, we observed the characteristic speedmeter behavior of the transfer function. The results are shown in Fig.~\ref{fig:4}, demonstrating clear difference between the position and 
speed frequency behaviour, as well as the good match with our theoretical model presented in the Appendix.

\begin{figure}
      \includegraphics[width=0.9\textwidth]{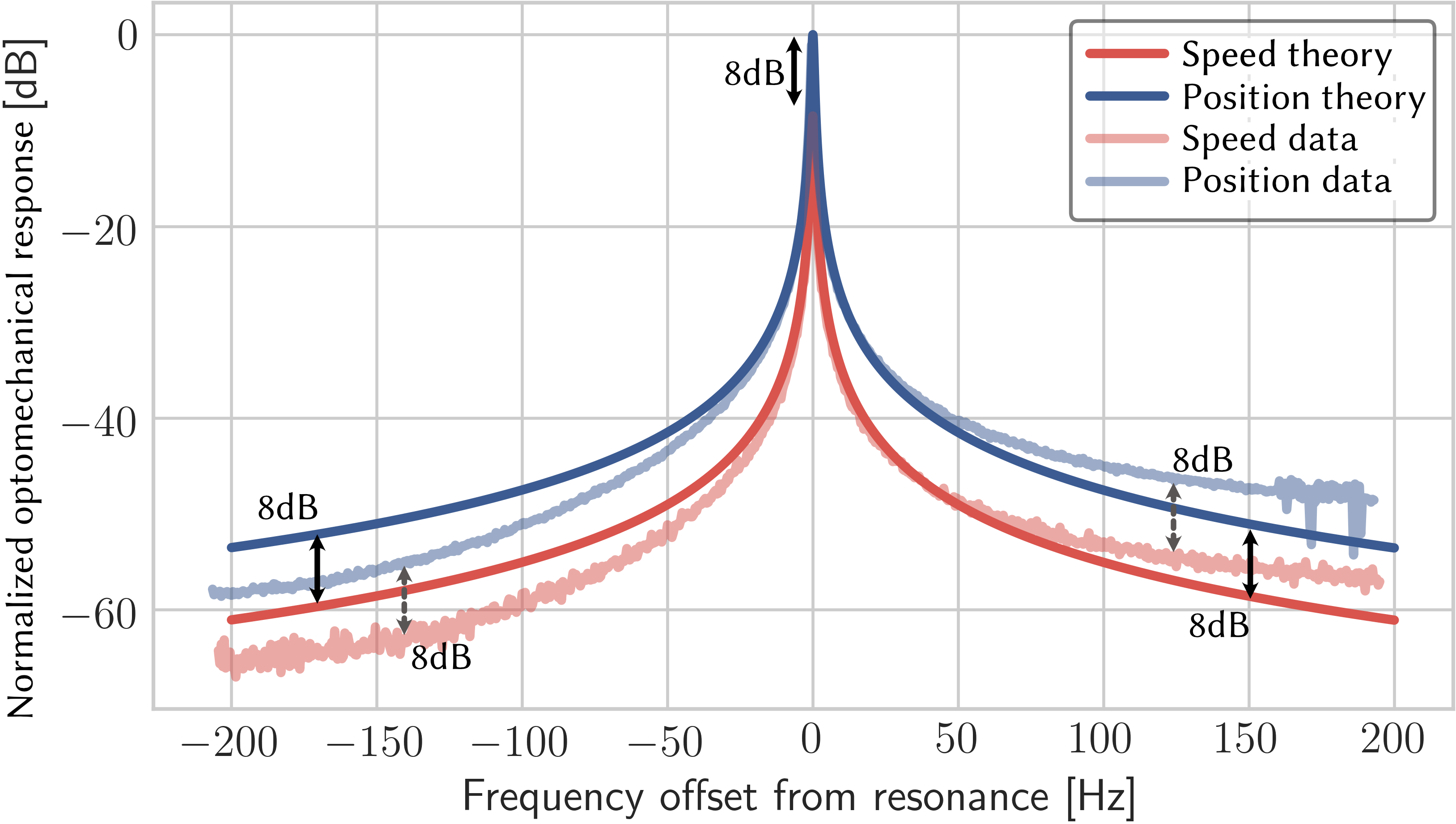}
  \caption{
  Optomechanical response of the membrane to an external force (PZT actuation). Shown are BHD measurements on the measurement beams from the two HRC output ports. While the frequency behaviours are indistinguishable, the signal magnitude differs by approximately 8\,dB over the observation band. 
    This value corresponds to the expected difference in signal transfer function $\sim \gamma^2/\Omega^2 \sim 8$\,dB, thus confirming the general expectation of the speed meter property of the HRC.
  Theoretical curves are not fitted to the data, but are predictions based on the independently measured parameters: membrane frequency 395.2\,kHz, membrane Q factor $4.6\times 10^5$, cavity linewidth 0.84\,MHz, membrane reflectivity 4.6\%.
  We attribute the differences in the frequency slopes to  unmodeled effects of low-frequency vibrational excitation from the holder of the membrane and electronic transfer functions.
  }
  \label{fig:5}
\end{figure}

The second signature of the speedmeter was observed in a force sensing experiment by comparing the two output ports of the HRC.
Despite the clear speedmeter optical response, sensing the speed of the membrane was not directly possible due to fundamental limitations of the system: close to the mechanical resonance, position and speed are not distinguishable in mechanical oscillators, since they change role over the oscillation period.
Speedmeter response might have been observed far above the mechanical resonance, where the membrane acts as a free mass. 
However, at those frequencies the effects of the high-order mechanical modes contaminated the speed signal, as is shown in Fig.\,\ref{fig:A8} of the Appendix.
Instead of the different frequency scaling of the position and speed signals, we observed the difference in strength of each signal, matching the expectations of Eq.\,\ref{eq:ring:transfer} and a rigorous calculation presented in the Appendix. 
We excited the membrane at different frequencies with a piezo actuator mounted to the membrane holder.
We then recorded the optical signal from its motion at position and speed ports correspondingly.
As expected, no difference in frequency scaling between the speed and position signals was visible, see\,Fig.~\ref{fig:5}.
Frequency scaling of the experimental data differed from the theory prediction due two effects. First, the transfer function of electronics and piezo actuators that were not included in the model and thus the excitation strength was not equal at all frequencies. Second, the low-frequency vibrations of the membrane frame coupled to the motion of the membrane, creating beat notes around the membrane peak (not visible in the figure) and applying additional frequency dependence to the response.\\

The outcome of the second experiment was that higher-order mechanical modes of the membrane prevented a direct observation of the characteristic speedmeter frequency response. However, since our HRC provides both a speed output and a position output, we were able to infer the speedmeter behavior by comparing the magnitudes of the otherwise similar signals. Replacing the membrane with a pendulum-suspended test mass should enable direct observation of the distinct speedmeter frequency response, including its full noise behavior.

To highlight the special features of the HRC with respect to speed and position measurements, we develop the full quantum-mechanical theory of the sensitivity for this setup.

\section{Theoretical analysis of the hybrid readout cavity speedmeter}
\subsection{Introduction to the speedmeter concept}
Before we analyse the HRC, we consider the general speed meter configuration, where the light bounces off a movable mirror from two sides, experiencing some delay between the bounces\,\cite{Danilishin2012}, as shown in Fig.\ref{fig:1}.
When assuming the mirrors to be perfectly reflective the input-output relations for the light field quadratures read as follows\,\cite{Caves1985a, Schumaker1985a, Danilishin2012}:

\twocolumequationpair{
&  \hat{b}^{c}(t) = \hat{a}^{c}(t),\\
&  \hat{b}^{s}(t) = \hat{a}^{s}(t) + 2ik_p A \left(x(t-\tau)-\hat{x}(t)\right),
}{eq:speed:inout}{
&  \hat{b}^{c}(t) = \hat{a}^{c}(t),\\
&  \hat{b}^{s}(t) = \hat{a}^{s}(t) - 2ik_p A \hat{x}(t),}{eq:position:inout}

\noindent where $\hat{a},\hat{b}^{c,s}$ are the input and output cosine and sine quadratures respectively, $k_p$ is the wave vector of the light field, $A$ is the amplitude of the light field, $\hat{x}(t)$ is the displacement of the mirror, and $\tau$ is the time delay between the two bounces.
In frequency domain, assuming the delay to be small, the input-output relations can be written as:

\twocolumequationpair{\hat{b}^{c}(\Omega) & = \hat{a}^{c}(\Omega),\\
  \hat{b}^{s}(\Omega) & = \hat{a}^{s}(\Omega) - 2 k_p A \hat{x}(\Omega) \left(1-e^{i\Omega\tau}\right)\nonumber\\
& \approx \hat{a}^{s}(\Omega) - 2k_p A \Omega\tau \hat{x}(\Omega).
}{eq:velocity:freq}{
  \hat{b}^{c}(\Omega) & = \hat{a}^{c}(\Omega),\\
  \hat{b}^{s}(\Omega) & = \hat{a}^{s}(\Omega) - 2 k_p A \hat{x}(\Omega)
}{eq:position:freq}

Here, one can already see the characteristic property of the speedmeter: the output contains the displacement proportional to the frequency $\Omega$, which is a consequence of the fact that speed and position are connected by a Fourier transform as: $v(t)=\dot{x}(t)\rightarrow -i\Omega x(\Omega)$.
Displacement $x(\Omega)$ has contribution from a signal force $G$, which we would like to measure, and radiation-pressure force $F_{\mathrm{rp}}$, which contaminates the measurement:
\begin{equation}
  \hat{x}(\Omega) = \chi(\Omega)(G(\Omega)+\hat{F}_{\mathrm{rp}}(\Omega)),
\end{equation}
where $\chi(\Omega)$ is the mechanical transfer function.
For the case of a free mass it is related to the mirror mass $M$ by $\chi(\Omega) = -(M\Omega^2)^{-1}$.

Radiation-pressure fluctuating force $\hat{F}_{\mathrm{rp}}(t) = \delta \hat{P}(t) c^{-1}$ depends on the fluctuations $\delta \hat{P}(t)$ in the power of the light field. For the speedmeter, the radiation-pressure force is given by the difference between the two bounces, which can be written as:

\twocolumequationpair{
&  \hat{F}_{\mathrm{rp}}(t) = 2\hbar k_p A \left(\hat{a}^{c}(t-\tau)-\hat{a}^{c}(t)\right),\\
&  \hat{F}_{\mathrm{rp}}(\Omega) \approx -2i\hbar k_p A \hat{a}^{c}(\Omega) \Omega \tau .}{eq:velocity:rp}{
&  \hat{F}_{\mathrm{rp}}(t) = - 2\hbar k_p A \hat{a}^{c}(t),\\
&  \hat{F}_{\mathrm{rp}}(\Omega) \approx -2 \hbar k_p A \hat{a}^{c}(\Omega).
}{eq:position:rp}
Here, the key difference between the speedmeter and the position meter is clear: in the speedmeter, the radiation-pressure force is proportional to the frequency, which leads to its cancellation at low frequencies, unlike the position meter.

The output light $\hat{b}$ is measured on a balanced homodyne detector, which allows to measure an arbitrary quadrature $\zeta$ of the light field.
The resulting signal is:
\begin{equation}
  \hat{y}(\Omega) = \hat{b}^{c}(\Omega) \cos\zeta + \hat{b}^{s}(\Omega)\sin\zeta,
\end{equation}
which results in the corresponding equations for the speedmeter and position meter:

\twocolumequationpair{
    \hat{y}(\Omega) & = \hat{a}^{c}(\Omega)\cos\zeta + \hat{a}^{s}(\Omega)\sin\zeta - \nonumber\\
   & - 2ik_p A \Omega\tau \chi(\Omega) \left(G(\Omega)- \right. \nonumber\\
   & \left.- 2i\hbar k_p A \hat{a}^{c}(\Omega) \Omega \tau \right)\sin\zeta.
}{eq:velocity:output}{
   \hat{y}(\Omega) & = \hat{a}^{c}(\Omega)\cos\zeta + \hat{a}^{s}(\Omega)\sin\zeta - \nonumber\\
   & - 2 k_p A \chi(\Omega) \left(G(\Omega)- \right. \nonumber\\
   & \left.- 2\hbar k_p A \hat{a}^{c}(\Omega) \right)\sin\zeta.
}{eq:position:output}

Since we are interested in the displacement of a test mass under the influence of the external force, we can normalize the output signal to the displacement, i.e. divide by the factor in front of $\chi(\Omega)G(\Omega) = x(\Omega)$:
\twocolumequationpair{\tilde{y}(\Omega) = -\hat{y}(\Omega) \frac{1}{2ik_p A \Omega\tau \sin\zeta}}{eq:velocity:norm}{
    \tilde{y}(\Omega) = -\hat{y}(\Omega) \frac{1}{2k_p A  \sin\zeta}
}{eq:position:norm}

To quantify the sensitivity, we define the spectral densities of the output signal\,\cite{Danilishin2012}:
the shot noise $S_{xx}$, the radiation-pressure noise $S_{FF}$ and their cross-correlation $S_{xF}$, assuming the input noise to be in a coherent state, i.e.\,$S_{a^{c}a^{c}} = S_{a^{s}a^{s}} = 1$, $S_{a^{c}a^{s}}=0$:
\twocolumequationpair{&S_{xx}(\Omega) = \frac{1}{4k_p^2 A^2 \Omega^2 \tau^2 \sin^2\zeta},\\
  &S_{FF}(\Omega) = 4\hbar^2 k_p^2 A^2 \Omega^2 \tau^2,\\
  &S_{xF}(\Omega) = -\hbar \cot \zeta.}{eq:velocity:sd}{
  &S_{xx}(\Omega) = \frac{1}{4k_p^2 A^2 \sin^2\zeta},\\
  &S_{FF}(\Omega) = 4\hbar^2 k_p^2 A^2,\\
  &S_{xF}(\Omega) = -\hbar \cot \zeta.
  }{eq:velocposition:sd}

The total noise spectral density normalized to displacement is\,\cite{Danilishin2012}:
\begin{equation}
  S_{x}(\Omega) = S_{xx} - \frac{2\Re[S_{xF}(\Omega)]}{M\Omega^2} + \frac{S_{FF}(\Omega)}{M^2\Omega^4},
\end{equation}
which results in the following expressions for the speedmeter and position meter sensitivity:

\twocolumequationpair{
  S_x(\Omega) = & \frac{\hbar}{M\Omega^2}\left(\frac{c M}{4k_p P \tau^2\sin^2\zeta} + 2\cot \zeta +\right. \nonumber \\ + & \left. \frac{4 k_p P \tau^2}{M c}\right),
  }{eq:velocity:sens}{
    S_x(\Omega) = & \frac{\hbar}{M\Omega^2}\left(\frac{c M\Omega^2}{4k_p P \sin^2\zeta} + 2\cot \zeta +\right.\nonumber \\  + & \left. \frac{4 k_p P}{M \Omega^2 c}\right),
  }{eq:position:sens}
\noindent where we defined the average optical power $P = \hbar k_p c A^2$.
We note here the key feature of the speedmeter: frequency dependence for all three contributions to the sensitivity is the same, i.e.\,the sensitivity is proportional to $1/\Omega^2$.
For a position meter, each contribution has its own frequency dependence.
In a standard case the homodyne detector measures the phase quadrature of the light: $\zeta = \pi/2$, and no cross-correlation between the noises is present:

\twocolumequationpair{& S_{x}(\Omega) = \frac{\hbar}{M\Omega^2}\left(\frac{c M}{4k_p P \tau^2} + \frac{4 k_p P \tau^2}{M c}\right).
}{eq:velocity:quad}{ S_{x}(\Omega) = \frac{\hbar}{M\Omega^2}\left(\frac{c M \Omega^2}{4k_p P} + \frac{4 k_p P}{M \Omega^2 c}\right).
}{eq:position:quad}
It is evident that the sensitivity of the speedmeter simply follows the frequency scaling of the SQL $S_{\rm SQL} = 2\hbar/M\Omega^2$, while the position meter only touches it at one specific frequency.
Notice the direct relation to Eqs.\,\ref{eq:general_sens},\ref{eq:position:scale}.
As discussed in the first section, a speedmeter has a significantly better sensitivity at low frequencies than a position meter, even if no quantum correlations are present.

The sensitivity of the detector can be further improved by employing optimal correlations between the SN and the QRPN, which can be achieved by optimizing the homodyne angle $\zeta$ (or, alternatively, injecting squeezed light in particular quadrature)\,\cite{Kimble2001,Danilishin2012}:
\twocolumequationpair{\cot \zeta = - \frac{4 k_p P \tau^2}{M c}}{eq:velocity:opt}{\cot \zeta (\Omega) = - \frac{4 k_p P}{M \Omega^2 c}.}{eq:position:opt}
\noindent In this case the sensitivity can surpass the SQL for both the speedmeter and the position meter:

\twocolumequationpair{S_{x}(\Omega) = \frac{\hbar c}{4k_p P \tau^2 \Omega^2}.
}{eq:velocity:subsql}{S_{x}(\Omega) = \frac{\hbar c}{4k_p P}.
}{eq:position:subsql}
\noindent Crucially, a position meter requires frequency-dependent angle, while a speedmeter achieves sub-SQL sensitivity by simply choosing the optimal frequency-\textit{independent} homodyne angle $\zeta$. This comes at a price of the reduced sensitivity by a factor of $\Omega^2 \tau^2$, but this difference significantly diminishes if optical losses are taken into account\,\cite{Danilishin2012}.
Similarly, speedmeter does not require frequency-dependent squeezing to achieve sub-SQL sensitivity, which allows to avoid using costly filter cavities.

Now, having compared the quantum noise a speedmeter and a position meter, we can discuss how this applies to the HRC.

\subsection{Speedmeter configuration}
Achieving the speedmeter sensitivity requires particular frequency dependence of all contributions to the sensitivity: SN, QRPN and their cross-correlation.
The first requires having the speedmeter transfer function for the signal, i.e. having it proportional to frequency.
The second means that QRPN should be proportional to frequency as well, for frequencies well below cavity linewidth $\Omega \ll \gamma$.
If both SN and QRPN have the appropriate scaling, so does their cross-correlation.
As we show in this section, the HRC naturally satisfies the first requirement, and the second requirement can be achieved through either adjusting the optical parameters of the setup, using particular quantum correlations or optimal post-processing of the output signals.

The speed port remains dark for the average classical field, i.e.\,the full power is reflected back to the position port.
Following the notation in Fig.\,\ref{fig:2} and defining average fields by a capital letter $\langle \hat{a}_i(t) \rangle = A_i$, the dark port condition can be expressed, as shown in the appendix (Eq.\,\ref{eq:17}, assuming the laser injection only from one side: $A_2 = 0$):
\begin{align}
&B_1 = A_1\\
&B_2 = 0
\end{align}
For the quantum fields, since the macroscopic position of the test mass does not affect the resonance of the system, we can set without loss of generality $L_1 = L_2 = L/2$.
We then obtain quantum amplitudes of the fields on the two ports:
\begin{align}\label{eq:ring:inout}
\hat{b}_1 = \beta_{11} \hat{a}_1 + \beta_{12} \hat{a}_2 +  \beta_{13} \hat{x}, \\
\hat{b}_2 =  \beta_{21} \hat{a}_1 +  \beta_{22} \hat{a}_2 +  \beta_{23} \hat{x},
\end{align}
where $\hat{x}$ is the displacement of the test mass, and the coefficients are computed from the full input-output formalism, presented in the Appendix:
\begin{align}\label{eq:ring:transfer}
& \beta_{11}(\Omega) = \beta_{22}(\Omega) \approx \frac{\gamma}{\gamma - i\Omega},\\
& \beta_{12}(\Omega)=  \beta_{21}(\Omega) \approx \frac{-i\Omega}{\gamma - i\Omega},\\
& \beta_{13}(\Omega) \approx 2 i k_p A_1 \left(1 + \frac{i\Omega}{2(\gamma - i\Omega)}\right),\\
& \beta_{23}(\Omega) \approx i\Omega\frac{2 i k_p A_1}{2(\gamma - i\Omega)}.
\end{align}
There are two important features to notice in these equations. First,the signal on the speed port $ \beta_{23}(\Omega)$ is proportional to frequency, as it should be for a speedmeter.
At low frequencies, where $\Omega\ll\gamma$, this condition is fully satisfied, and we have a pure speed signal.
At high frequencies, the cavity bandwidth averages the signal, which turns the signal into a position one (i.e.\,zeroth order in frequency): $ \beta_{23}(\Omega\gg\gamma)\approx -ik_pA_1$.
The position port contains a mixture of the position and the speed signal, since $ \beta_{13}$ contains both zeroth and first order in frequency,
approaching purely position contribution for small and large frequencies (and only getting the mixture of speed in the intermediate regime $\Omega\sim\gamma$).

The second important point to be seen in the equations is the optical transfer functions for the incoming fields $ \beta_{11,12,21,22}$.
The noise coupling from one port to another acquires a speedmeter scaling ($\beta_{12, 21}$), and the noise reflected directly to the position port does not($\beta_{11, 22}$), similarly to the behavior of the average amplitudes. Once we normalize the output shot noise to the corresponding signal transfer functions $\beta_{13, 23}$, we obtain the displacement spectral densities:
\begin{align}
&S_{x, 1} = \frac{\hbar c^2}{4 I_{\mathrm{in}} \omega_p} \frac{\gamma^2 + \Omega^2}{\gamma^2 + \Omega^2/4},\\
&S_{x, 2} = \frac{\hbar c^2}{4 I_{\mathrm{in}} \omega_p} \frac{\gamma^2 + \Omega^2}{\Omega^2}.
\end{align}
Before we proceed to compute the radiation pressure noise contribution, it is instructional to compare the shot-noise-limited sensitivity of the HRC with the one of a standard position meter and a standard speedmeter.
{\begin{table}
\begin{center}
\begin{tabular}{l|c|c}
  \hline
  \hline
   & speedmeter & position meter \\
  \hline
  \rule{0ex}{4ex} without cavity & $\displaystyle\frac{\hbar c^2}{4 I_{\mathrm{in}}\omega_p} \frac{1}{ \Omega^2 \tau^2}$ & $\displaystyle\frac{\hbar c^2}{4 I_{\mathrm{in}} \omega_p}$ \rule[-2.5ex]{0ex}{4ex}\\
  \hline
  \rule{0pt}{4ex} HRC & $\displaystyle\frac{\hbar c^2}{4 I_{\mathrm{in}} \omega_p} \frac{T^4}{\Omega^2\tau^2}$ & $\displaystyle\approx\frac{\hbar c^2}{4 I_{\mathrm{in}} \omega_p}$ \rule[-2.5ex]{0ex}{4ex}\\
  \hline
  \rule{0pt}{4ex} standard detector & $\displaystyle\frac{\hbar c^2}{4 I_{\mathrm{in}} \omega_p} \frac{T^4}{4}\frac{T^4}{16 \Omega^2\tau^2}$ & $\displaystyle\frac{\hbar c^2}{4 I_{\mathrm{in}} \omega_p} \frac{T^4}{4}$ \rule[-2.5ex]{0ex}{4ex}\\
  \hline
  \hline
\end{tabular}\vskip7pt\caption[Comparison between the shot-noise-limited sensitivities of different detectors.]{Comparison between the shot-noise-limited sensitivities of different detectors.
The columns represent speed or position measurement in different detector topologies, represented by the rows.
The rows define the type of the detector: position and speedmeters without optical cavities, two output ports of the HRC, and a standard Michelson-Fabry-Perot position meter and a Sagnac speedmeter. }\label{tab:ring:x_vs_v}
\end{center}
\end{table}
}
In Table\,\ref{tab:ring:x_vs_v} we summarize the main results of the comparison.
In a position meter, an optical cavity enhances the optical power and the signal through constructive interference, which results in sensitivity enhancement proportional to the square of the cavity finesse, i.e. $\sim T^4$.
In a standard speedmeter, optical cavities play two fundamental roles: they enhance the light power and signal, but also increase the effective delay $\tau \rightarrow \tau/T^2$, which benefits the sensitivity of a speedmeter.
In the HRC speedmeter, the optical cavity plays a different role. 
Due to destructive interference between the multiple round trips of the light field, the signal is not enhanced by the cavity finesse.
At the same time, the effective delay is enhanced by the cavity finesse, which results in a speed sensitivity that is $\sim T^4$ times better than the one of a free mirror without cavity.
Position sensitivity remains the same as for a case without any cavity.

The HRC behavior is thus rather counter-intuitive: the light power inside the cavity is enhanced by the cavity finesse, but the signal does not benefit from this enhancement.
The reason for this is the destructive interference between the signal that is reflected off the test mass mirror and the same signal on one round trip inside the cavity. 
Only a small portion of the signal is not canceled by this interference, and leaks out through the front mirror, proportionally to the incoming light power, in accordance with energy conservation.
It is possible to enhance the signal by implementing an additional cavity around the test mass. 
Such configuration would achieve the traditional signal strength of a cavity-enhanced system.

The key feature of a speedmeter is its reduced QRPN, which is achieved by the destructive interference between the quantum radiation pressure force from the two sides of the test mass.
Such cancellation would result in a pure speed scaling of the radiation-pressure noise.
In the HRC, this cancellation is achieved up to a small power imbalance between the light fields traveling clock-wise and counter-clockwise due to the portion leaking out through the front mirror.
As a result, radiation pressure force has both the speedmeter contribution, and the position-meter contribution (see Appendix for more details):
\begin{multline}\label{eq:ring:frp1}
  \hat{F}_{\mathrm{rp}}(\Omega) = \frac{\hbar k_p A}{\sqrt{2} (\gamma-i\Omega)} \left[ \left(r^2 - t^2\right) \left(2 \gamma - i\Omega\right)\hat{a}^c_1(\Omega) + i rt\Omega \hat{a}^s_1(\Omega) + \right. \\ + \left. \left(r^2 - t^2\right) \Omega \hat{a}^c_2(\Omega)  + i r t\left(2 \gamma - i\Omega\right) \hat{a}^s_2(\Omega)\right].
\end{multline}
We can see that the force has both position (zeroth order in frequency) and speed (first order in frequency) contributions.
This could also be understood from the quantum measurement theory perspective: each measurement produces some back-action on the measured system (unless it is a QND measurement), and since the HRC system measures some position signal, it must produce some back-action on the position of the test mass, which is the radiation pressure noise.
If we assume the test mass to be perfectly reflective, $r=1$, $t=0$, the spectral density of radiation pressure force becomes:
\begin{equation}
S_{\mathrm{rp}}(\Omega) = 2\gamma^2\frac{\hbar \omega_p I_{\mathrm{in}}}{c^2 (\gamma^2 + \Omega^2)} +  \Omega^2\frac{\hbar \omega_p I_{\mathrm{in}}}{c^2 (\gamma^2 + \Omega^2)},
\end{equation}
where $I_{\mathrm{in}} = \frac{1}{2}\hbar k_p c A^2$ is the average optical power in the incoming beam.
In order to suppress the position contribution to the radiation pressure and achieve a purely speedmeter scaling, we could entangle the two input (quantum) fields such, that
$(r^2-t^2)\hat{a}_1^c = -irt \hat{a}_2^s$, and the position-dependent part in Eq.\,\ref{eq:ring:frp1} gets canceled.
If the mirror is semi-transparent, $r=t$, the only position contribution comes from $a_2^s$, which can be suppressed by injecting squeezed light into the second port.
In a general case, position contribution can be suppressed by measuring both output ports and optimally combining the outputs, as we show in the Appendix.

While we do not investigate in details the applicability of the HRC for a large-scale detector design, we point to the work in Refs.\,\cite{guoMergingLshapedResonator2024, guoSensingControlScheme2023, zhangGravitationalWaveDetectorPostmerger2023a} that essentially realizes the HRC with a different motivation of exploiting the specific resonance features and shows that such a design can achieve the sensitivity comparable to a standard detector. Speedmeter features are not investigated there and will be the topic of future studies, as we briefly discuss in the Appendix.

\section{Conclusion}
An interferometer built as a speedmeter measures the differential change in speed of mirrors. 
Its design differs significantly from a conventional interferometer, which measures the differential change in position of mirrors.
A speedmeter for the detection of GWs that is enhanced with injected squeezed light would achieve a sensitivity better than the SQL without long-baseline narrow-linewidth filter cavities. Neither a frequency-dependent squeeze angle nor a variational readout would be needed, which is different from a conventional interferometer\,\cite{Kimble2001}.
On the other hand, the realization of a speed meter presents new challenges. 
Many of these have not yet been investigated experimentally. 
With this work we present an interferometer, a ``hybrid readout cavity'', which provides two optical outputs where the performance of speedmeters can be directly compared with those of the conventional position meter relative to each other. 
The cavity combines the traveling wave of a ring resonator with the standing wave of a Fabry-Perot resonator. The optical coupling is provided by a moving semi-transparent mirror.
In our experiments, the moving mirror was a SiN membrane.
It became apparent during the project that the higher order membrane vibration modes make some speedmeter properties unmeasurable. Others could be measured here. 
Future research on the ``hybrid readout cavity'' should use a mirror suspended as a pendulum so that all speedmeter properties can be measured in comparison to the position meter.\\
Our results highlight the need for further research and development in the area of quantum speedmeters, experimental studies of their unusual behavior, both for gravitational wave detection and for other advanced sensing applications.
\
\section*{Acknowledgements}
While preparing this manuscript, we became aware of related work demonstrating the speedmeter transfer function of a table-top setup, which is now being published independently\,\cite{kranzhoff2025}. We acknowledge the complementary nature of both efforts and believe they contribute jointly to the development of the field.

The authors are grateful to Farid Khalili for fruitful discussions.
The work of MK and RS was supported by the Deutsche Forschungsgemeinschaft (DFG) (No. SCHN 757/6-1) and by the Deutsche Forschungsgemeinschaft (DFG) under Germany's Excellence Strategy EXC 2121  ``Quantum Universe''-390833306. Y. M. is supported by the National Key R\&D Program of China (2023YFC2205801), National Natural Science Foundation of China under Grant No.12474481.

\section*{References}
\bibliography{RSbib, MKbib}

\vspace{10mm}
{\Large \bf Appendix}
\section{Appendix: full theoretical model}
\subsection{Input-output relations}
We start the derivation of the quantum-limited sensitivity by solving the input-output relations for the optical fields in the cavity in matrix form, and then study the properties of the solutions.
We start define the propagation matrices for light fields in frequency domain, see Fig.\,\ref{fig:2}:
\begin{align}
&\mathbb{R} = \begin{pmatrix}
R & 0\\ 0 & R
\end{pmatrix} \qquad  \text{reflection matrix;} \\
&\mathbb{T} = \begin{pmatrix}
T & 0\\ 0 & T
\end{pmatrix} \qquad  \text{transmission matrix;}\\
&\mathbb{M} = \begin{pmatrix}
r & it \\ it & r
\end{pmatrix} \qquad \text{reflection/transmission of the test mass;}  \\
&\mathbb{P}(\Omega) = \begin{pmatrix}
e^{i(k_p+\Omega/c)L_1} & 0 \\ 0 & e^{i(k_p+\Omega/c)L_2}
\end{pmatrix} \qquad \text{propagation of quantum fields;}\\
& \mathbb{P}(0) = \begin{pmatrix}
e^{ik_pL_1} & 0 \\ 0 & e^{ik_pL_2}
\end{pmatrix} \qquad \text{propagation of semi-classical fields.}
\end{align}
Here $k_p$ is the laser field wave vector, $R$ is the amplitude reflectivity of the front mirror (we assume the other mirrors perfectly reflective), $T$ is the amplitude transmissivity of the front mirror, $r, t$ are the amplitude reflectivity and transmissivity of the test mass, $L_{1,2}$ are the distances from the front mirror to the test mass (clock and counterclockwise), $\Omega$ is the sideband Fourier frequency and $c$ is the speed of light.
In addition, we define some useful matrices:
\begin{equation}
\sigma_1 = \begin{pmatrix}
0 & 1 \\ 1 & 0
\end{pmatrix}, \quad
\sigma_2 = \begin{pmatrix}
0 & -i \\ i & 0
\end{pmatrix}, \quad
\sigma_3 = \begin{pmatrix}
1 & 0 \\ 0 & -1
\end{pmatrix}, \quad
\mathbb{I} = \begin{pmatrix}
1 & 0 \\ 0 & 1
\end{pmatrix}
\end{equation}

\subsubsection{Classical field}
First, it is useful to write the equations of motion for the average semi-classical fields (i.e.\,at frequency $\Omega=0$).
We combine two propagating fields into a single vector, represented by a bold letter, e.g. $$\mathbf{B} = \begin{pmatrix}
B_1 \\ B_2
\end{pmatrix}$$
In this notation the input-output equations are:
\begin{align}\label{eq:inout}
&\mathbf{B} = -\mathbb{R} \sigma_1 \mathbf{A} + \mathbb{T}\mathbf{C},\\
&\mathbf{C} = \mathbb{P}(0)\mathbf{F},\\
&\mathbf{F} = \mathbb{M}\mathbf{E},\\
&\mathbf{E} = \mathbb{P}(0)\mathbf{D},\\
&\mathbf{D} = \mathbb{R}\sigma_1\mathbf{C} + \mathbb{T}\mathbf{A}.
\end{align}
This set of equations can be written for the intra-cavity field $\mathbf{C}$:
\begin{equation}
\mathbf{C} = \mathbb{P}(0)\mathbb{M}\mathbb{P}(0)\left(\mathbb{R}\sigma_1\mathbf{C}+\mathbb{T}\mathbf{A}\right),
\end{equation}
from which we obtain the input-output relation for the intra-cavity field:
\begin{align}
&\mathbf{C} = \left(\mathbb{I}-\mathbb{P}(0)\mathbb{M}\mathbb{P}(0)\mathbb{R}\sigma_1\right)^{-1}\mathbb{P}(0)\mathbb{M}\mathbb{P}(0)\mathbb{T}\mathbf{A} = \mathbb{K}(0)\mathbb{P}(0)\mathbb{M}\mathbb{P}(0)\mathbb{T}\mathbf{A},\\
&\mathbb{K}(0) = \left(\mathbb{I}-\mathbb{P}(0)\mathbb{M}\mathbb{P}(0)\mathbb{R}\sigma_1\right)^{-1}.
\end{align}
Substituting back into Eqs.\,\ref{eq:inout}, we obtain the solutions for the outgoing field $\mathbf{B}$ and field at the test mass $\mathbf{E}$:
\begin{align}
\mathbf{B} &= \left(-\mathbb{R}\sigma_1 + \mathbb{T}\mathbb{K}(0)\mathbb{P}(0)\mathbb{M}\mathbb{P}(0)\mathbb{T}\right)\mathbf{A},\\
\mathbf{E} &= \mathbb{P}(0)\left(\mathbb{R}\sigma_1\mathbb{K}(0)\mathbb{P}(0)\mathbb{M}\mathbb{P}(0) + \mathbb{I}\right)\mathbb{T}\mathbf{A}.
\end{align}
We keep these expressions in the general form for the time being, writing down explicit expressions in the approximated solution in the next section.

\subsection{Normal mode splitting of the hybrid readout cavity} \label{ssec:resonances}
\begin{figure}
  \begin{minipage}{1\textwidth}
    \includegraphics[width=1\textwidth]{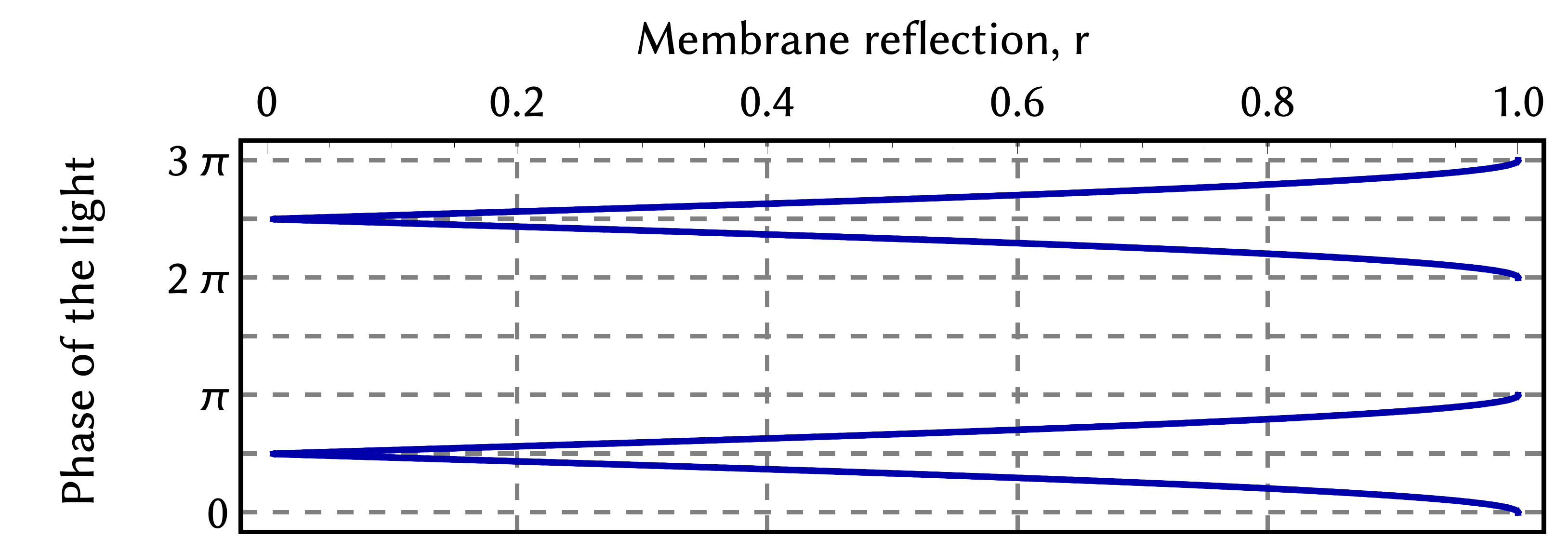}
  \end{minipage}
  \vspace{-0.2cm}
  \begin{minipage}{1\textwidth}
    \includegraphics[width=1\textwidth]{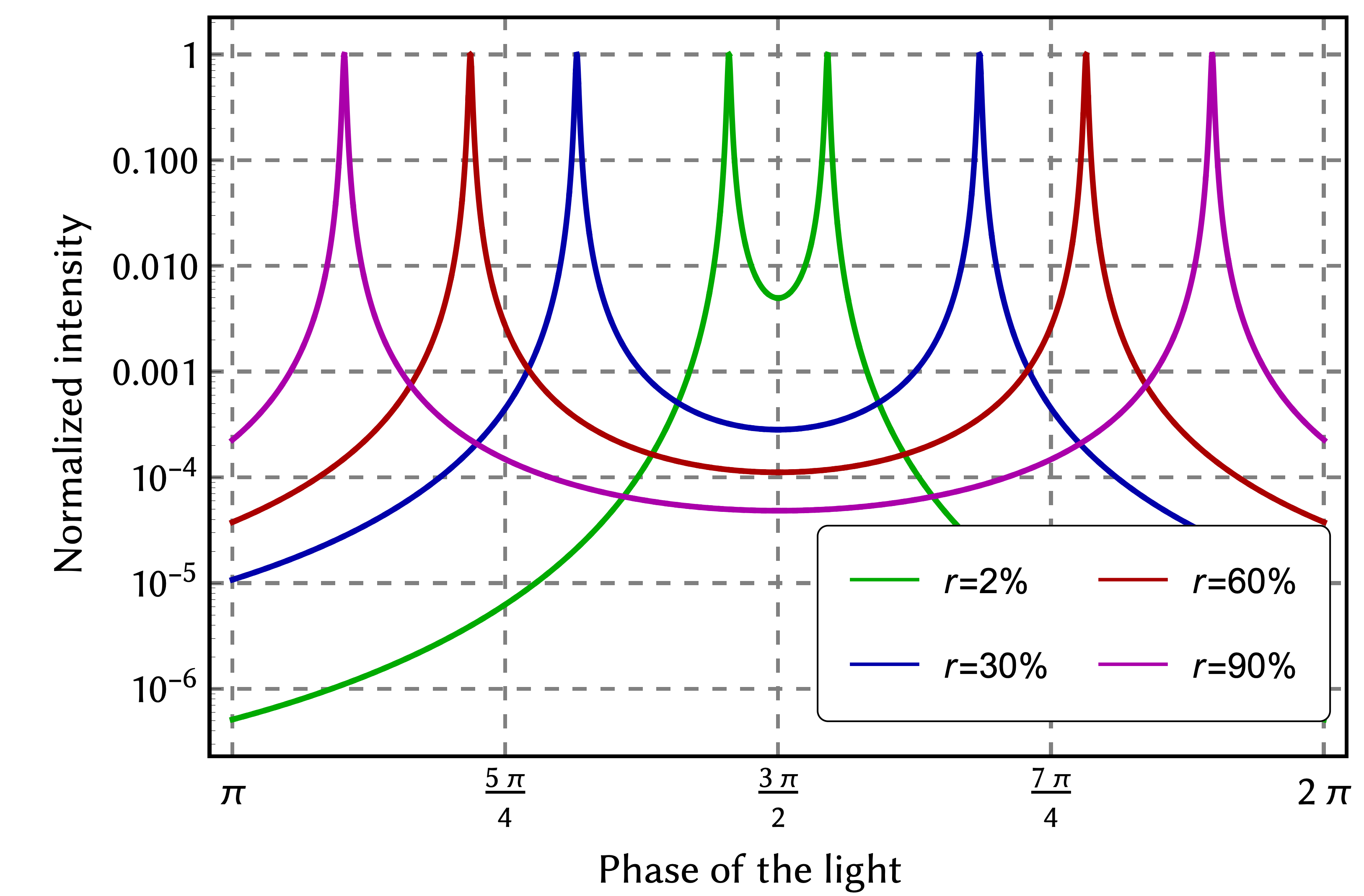}
  \end{minipage}
  \caption{Resonance structure of the HRC.
  Top: the resonance condition for the phase of the incoming light as a function of membrane reflection coefficient.
  Bottom: intensity of intra-cavity field normalized to its peak as a function of the phase of incoming light for different reflectivities of the membrane.
  Without a membrane, $r=0$, the peaks are separated by one FSR ($2\pi$ phase), which corresponds to the empty ring cavity.
  With a membrane the resonance splits into two, and the separation between the two modes increases with increased reflectivity.
  When the membrane is perfectly reflective, the peaks are equidistant again, corresponding to the Fabry-Perot cavity case.}
  \label{fig:ring:resonances}
\end{figure}

Using these expressions, we study the resonance condition in the HRC. 
We compute the average amplitudes of the output fields on the two ports of the cavity as a function of the incoming field amplitude $A_{1,2}$:
\begin{equation}
B_{1,2} = \frac{A_{2,1}}{\mathcal{D}(0)} \left[R \left(1-e^{2 i k_p L}\right)-i t \left(R^2+1\right) e^{i k_p L}\right]- \frac{A_{1,2}}{\mathcal{D}(0)} r T^2 e^{2 i k_p L_{1,2}} \, ,
\label{eq:17}
\end{equation}
where $L=L_1 + L_2$ is the HRC round trip length, $R, T$ are the input mirror amplitude reflectivity and transmissivity, $r, t$ are the test mass reflectivity and transmissivity, and $\mathcal{D}(\Omega)$ is the cavity resonance factor, the inverse of which defines the cavity amplification factor for each signal frequency $\Omega$:
\begin{equation}
\mathcal{D}(\Omega) = R^2  e^{2 i (k_p+\Omega/c) L}+2 i R t e^{i (k_p+\Omega/c) L}-1 \, .
\end{equation}
The resonance condition for the classical field corresponds to maximal enhancement of the intra-cavity field.
This can be found as the minimum of the resonance factor $\mathcal{D}(0)$ where the amplification inside the cavity is maximal on resonance. 
\begin{equation}
\frac{d|\mathcal{D}(0)|^2}{dk_p} = 0 \, ,
\end{equation}
which gives the following resonance conditions for the frequency $\omega_c = k_p c$
\begin{equation}
  \omega_c = \left\{
\begin{array}{l}
\frac{c}{L}\arcsin \left( -t\frac{1+R^2}{2R}\right) \\
\frac{c}{L}\left(\pi-\arcsin \left( -t\frac{1+R^2}{2R}\right)\right) \, .
\end{array}\right.
\end{equation}
Notice that the HRC resonances feature the normal mode splitting due to the non-zero test mass reflectivity, see Fig.\,\ref{fig:ring:resonances}

The unique and unusual property of the HRC is that the resonance condition does not depend on the position of the test mass, as it is the case for semi-transparent membranes in standing-wave cavities\,\cite{Thompson2008}.
Consider the case of a perfectly reflective test mass, $r=1$.
On resonance, a standing wave is formed, much as in a linear Fabry-Perot cavity.
However, the change in the test mass position does not cause the change of total length of the cavity, unlike the Fabry-Perot cavity.
Therefore, the resonance condition for the standing wave does not change either.
This fundamental property of the HRC leads to the emergence of the coherent optomechanical coupling\,\cite{Li2019}.
When the test mass's reflectivity is equal to zero, the resonance condition reproduces a usual ring cavity resonance with FSR $c/L$.
Such a cavity maintains two possible modes: clockwise and counter-clockwise, which have the same resonant frequency -- they are degenerate.
Introducing a reflecting element couples clockwise and counter-clockwise modes, and the interference between them results in two new resonant modes (symmetric and anti-symmetric).
These two new modes have different resonance frequencies, thus the test mass reflectivity breaks the degeneracy of the ring cavity.

When the reflectivity of the input mirror is rather high, the resonance condition can be approximated as:
\begin{align}
  &\cos k_{\pm} L \approx \pm r, \quad r = \cos \theta, \quad t = \sin \theta\\
  &\Delta \omega_c \equiv c(k_{+}-k_{-}) \approx \frac{2c}{L}\arcsin r \, .
\end{align}
Such normal mode splitting occurs also in other optomechanical systems, such as membrane-in-the-middle setups\,\cite{Jayich2008,Thompson2008,Peterson2015,Piergentili2018,Chen2020a}.
The unusual property in the HRC is that the splitting depends only on the reflectivity of the test mass, not its macroscopic position.

\subsubsection{Quantum field}
For the quantum field, it's necessary to keep the frequency dependence, and also consider that upon reflection off the test mass, the field picks up a phase modulation signal:
$$r \rightarrow re^{2ik x} \approx r (1 + 2 i k_p x(\Omega)),$$
where we expanded the exponent into a series for small $x$, and also assumed $k = k = k_p + \Omega/c \approx k_p$ for $\Omega \ll k_p c$.

Quantum fields are considered as a first order perturbation on top of the average semi-classical field amplitude: $\hat{A}(\Omega) = A + \hat{a}(\Omega)$.
The field that picks up the signal upon reflection is approximated to the first order of smallness: $r\hat{E}(\Omega)e^{2ik(\Omega) \hat{x}(\Omega)}\approx r E + r\hat{e}(\Omega) + 2ik_p r E \hat{x}$.
With these approximations, we write down the input-output relations for the quantum fields, omitting the hats over the quantum field vectors for brevity:
\begin{align}
&\mathbf{b}(\Omega) = -\mathbb{R} \sigma_1 \mathbf{a}(\Omega) + \mathbb{T}\mathbf{c}(\Omega),\\
&\mathbf{c}(\Omega) = \mathbb{P}(\Omega)\mathbf{f}(\Omega),\\
&\mathbf{f}(\Omega) = \mathbb{M}\mathbf{e}(\Omega) + 2ik_p x(\Omega) r \sigma_3 \mathbf{E},\\
&\mathbf{e}(\Omega) = \mathbb{P}(\Omega)\mathbf{d}(\Omega),\\
&\mathbf{d}(\Omega) = \mathbb{R}\sigma_1\mathbf{c}(\Omega) + \mathbb{T}\mathbf{a}(\Omega).
\end{align}
The solution to these equations for the intra-cavity field is:
\begin{align}
&\mathbf{c}(\Omega) = \mathbb{K}(\Omega)\mathbb{P}(\Omega)\mathbb{M}\mathbb{P}(\Omega)\mathbb{T}\mathbf{a}(\Omega) + 2ik_p x(\Omega) r \mathbb{K}(\Omega)\mathbb{P}(\Omega)\sigma_3 \mathbf{E},\\
&\mathbb{K}(\Omega) = \left[\mathbb{I}-\mathbb{P}(\Omega)\mathbb{M}\mathbb{P}(\Omega)\mathbb{R}\sigma_1\right]^{-1}.
\end{align}

The output field vector can also be obtained by solving the equations above:
\begin{align}
&\mathbf{b}(\Omega) = \left[-\mathbb{R}\sigma_1 + \mathbb{T}\mathbb{K}(\Omega)\mathbb{P}(\Omega)\mathbb{M}\mathbb{P}(\Omega)\mathbb{T}\right]\mathbf{a}(\Omega) + 2ik_p x(\Omega) r \mathbb{Y}(\Omega)\mathbf{A},\\
&\mathbb{Y}(\Omega) = \mathbb{T}\mathbb{K}(\Omega)\mathbb{P}(\Omega)\sigma_3\mathbb{P}(0)\left(\mathbb{R}\sigma_1\mathbb{K}(0)\mathbb{P}(0)\mathbb{M}\mathbb{P}(0) + \mathbb{I}\right)\mathbb{T}.
\end{align}

These equations are admittedly not convenient for studying the properties of the setup analytically, yet useful for generalized description and computer simulation.
For convenience, we can write the output fields on the two ports as:
\begin{align}\label{eq:ring:inout}
b_1 = \beta_{11} a_1 + \beta_{12} a_2 + \beta_{13} x, \\
b_2 = \beta_{21} a_1 + \beta_{22} a_2 + \beta_{23} x,
\end{align}
where the coefficients can be computed under the assumption of the bright field only injected from one side, i.e. $A_2 = 0$:
\begin{align}\label{eq:fullcoeff}
\beta_{11} = &\,\beta_{22} = -\frac{1}{\mathcal{D}(\Omega)} r T^2 e^{i (k_p + \Omega/c)L},\\
\beta_{12} = &\,b_{21} = -\frac{1}{\mathcal{D}(\Omega)} \left( -R + it (1+R^2)e^{i (k_p + \Omega/c)L} + R e^{2 i (k_p + \Omega/c)L}  \right),\\
\beta_{13} = &\,\frac{2ik_p r T^2 A_1 e^{i (k_p + \Omega/2c)L}}{\mathcal{D}(0)\mathcal{D}(\Omega)} \times \nonumber \\ &\times \left( 1 - R e^{i k_p L} (it + it e^{i \Omega L/c} + R e^{i (k_p + \Omega/c)L})\right),\\
\beta_{23} = &\,\frac{2ik_p A_1 r^2 T^2 R e^{i (2 k_p + \Omega/2c)L} \left(-1 + e^{i \Omega L/c} \right)}{\mathcal{D}(0)\mathcal{D}(\Omega)},\\
\mathcal{D}(\Omega) &= -1 + 2i t R e^{-i\theta} \left(1+ i \frac{\Omega L}{c}\right) + R^2 e^{-2i\theta} \left(1+ 2 i \frac{\Omega L}{c}\right).
\end{align}
The speedmeter behavior is not obvious in these expressions, but we can further simplify the expressions to make it more apparent.

To achieve that, we apply a single-mode approximation\,\cite{Danilishin2012} assuming:
\textit{(i)} the sideband (signal) frequency is much smaller than the cavity free spectral range $\Omega\ll c/L$, and
\textit{(ii)} the transmissivity $T$ of the front mirror is small, so that we can make a Taylor expansion $R \approx 1-T^2/2$.
We define the optical linewidth of the cavity as:
\begin{equation}
  \gamma = \frac{cT^2}{2L} = \frac{T^2}{2\tau},
\end{equation}
where $\tau = L/c$ is the round-trip time.
We can then rewrite the resonance factor in the signle-mode approximation:
\begin{align}
\mathcal{D}(\Omega) &= -1 + 2i t R e^{-i\theta} \left(1+ i \frac{\Omega L}{c}\right) + R^2 e^{-2i\theta} \left(1+ 2 i \frac{\Omega L}{c}\right) = \nonumber \\&= -2 r \tau e^{-i\theta} (\gamma - i \Omega),
\end{align}
where we applied the resonance condition:
$$e^{ik_pL} = r-it = e^{-i\theta}.$$
We then simplify the coefficients in Eqs.\ref{eq:fullcoeff} by going into the single-mode approximation:
\begin{align}\label{eq:ring:transfer:appendix}
&\beta_{11}(\Omega) = \beta_{22}(\Omega) \approx \frac{\gamma}{\gamma - i\Omega},\\
&\beta_{12}(\Omega)= \beta_{21}(\Omega) \approx \frac{-i\Omega}{\gamma - i\Omega},\\
&\beta_{13}(\Omega) \approx 2 i k_p A_1 \left(1 + \frac{i\Omega}{2(\gamma - i\Omega)}\right),\\
&\beta_{23}(\Omega) \approx i\Omega\frac{2 i k_p A_1}{2(\gamma - i\Omega)}.
\end{align}
This approximation can further be used to compute a concise form for spectral densities of the noises in the detector, which we do in the next section.

\subsubsection{Two-photon quadratures.}
In order to conveniently compute the spectral densities as measured with the homodyne detector, we transition to the two-photon quadratures\,\cite{Caves1985a,Schumaker1985a,Danilishin2012}.
Two-photon quadratures for a field $a(\omega_p \pm \Omega)$ are defined as following:
$$a^c(\Omega) = \frac{a(\omega_p + \Omega) + a^\dagger(\omega_p - \Omega)}{\sqrt{2}}, \qquad a^s(\Omega) = \frac{a(\omega_p + \Omega) - a^\dagger(\omega_p - \Omega)}{i\sqrt{2}}$$
and the corresponding vector:
$$\mathcal{a} = \begin{bmatrix}
a^c \\ a^s
\end{bmatrix}.
$$
Expressed in terms of two-photon quadratures, the input-output relations in Eq.\,\ref{eq:ring:inout} become:
\begin{align}\label{eq:ring:out2}
&\mathcal{b}_1 = \mathcal{R}_1(\Omega)\mathcal{a}_1 + \mathcal{R}_2(\Omega)\mathcal{a}_2 + x \mathcal{M}_{1x}(\Omega)\mathcal{A}_1 + \Omega x \mathcal{M}_{1v}(\Omega)\mathcal{A}_1,\\
&\mathcal{b}_2 = \mathcal{R}_2(\Omega)\mathcal{a}_1 + \mathcal{R}_1(\Omega)\mathcal{a}_2 + \Omega x \mathcal{M}_{2v}(\Omega)\mathcal{A}_1,
\end{align}
where the transfer matrices are:
\begin{align}
&\mathcal{R}_1(\Omega) = \frac{\gamma}{\gamma-i\Omega}\mathbb{I},\\
&\mathcal{R}_2(\Omega) = \frac{-i\Omega}{\gamma-i\Omega}\mathbb{I},\\
&\mathcal{M}_{1x}(\Omega) = -\frac{i k_p \gamma}{\gamma - i \Omega} \sigma_2,\\
&\mathcal{M}_{1v}(\Omega) = -\frac{i k_p}{2 (\gamma - i \Omega)}\sigma_2,\\
&\mathcal{M}_{2v}(\Omega) = \frac{k_p}{\gamma - i \Omega} \sigma_2.
\end{align}

\subsubsection{Shot noise spectral density}
The  displacement signal from the HRC is measured by a homodyne detector with homodyne phase $\zeta$:
\begin{equation}\label{eq:ring:outputshot}
  y_{1,2} = \mathbf{H}_{1,2}^T \mathcal{b}_{1,2}, \quad \mathbf{H}_{1,2}^T = \left[\cos\zeta, \sin\zeta\right]^{T}.
\end{equation}
In order to get the sensitivity, we normalize both outputs to the corresponding signal transfer function.
The way to do it can be dividing by the pre-factor in front of $x$ in Eqs.\,\ref{eq:ring:outputshot}:
\begin{align}
&\tilde{y}_1 = \frac{y_1}{\mathbf{H}_1^T \mathcal{M}_{1x} \mathcal{A}_1 + \Omega \mathbf{H}_1^T \mathcal{M}_{1v} \mathcal{A}_1} = \mathcal{X}_1 + x,\\
&\tilde{y}_2 = \frac{y_2}{\Omega \mathbf{H}_2^T \mathcal{M}_{2v} \mathcal{A}_1} = \mathcal{X}_2 + x,
\end{align}
where $\mathcal{X}_{1,2}$ are the quantum noise parts of the normalized output.
The signal is in a phase quadrature for the cavity tuned on resonance with incoming field, therefore we select $\zeta=\pi/2$, and also without loss of generality fix the phase of the incoming light such that
\begin{equation}
\mathcal{A}_1 = \sqrt{2}A
\begin{bmatrix}
1 \\ 0
\end{bmatrix}
= \sqrt{\frac{2 I_{\mathrm{in}}}{\hbar \omega_p}} \begin{bmatrix}
1 \\ 0
\end{bmatrix},
\end{equation}
where $I_{\mathrm{in}} = \frac{1}{2}\hbar k_p c A^2$ is the average optical power in the incoming beam.
With these notations we arrive at spectral densities of noises $\mathcal{X}_{1,2}$:
\begin{align}
&S_{\mathcal{X}_1}(\Omega) = \frac{\hbar c^2}{4 I_{\mathrm{in}} \omega_p} \frac{\gamma^2 + \Omega^2}{\gamma^2 + \Omega^2/4},\\
&S_{\mathcal{X}_2}(\Omega) = \frac{\hbar c^2}{4 I_{\mathrm{in}} \omega_p} \frac{\gamma^2 + \Omega^2}{\Omega^2}.
\end{align}
The speed output $\mathcal{X}_2$ evidently has the speedmeter scaling at low frequencies $\Omega \ll \gamma$.

\subsection{Radiation-pressure noise}
The second contribution to the quantum-noise-limited sensitivity of the detector is QRPN.
Each light field reflected off a test mass applies some force on it in the direction it impinges on it.
In the HRC, there are two sides of the test mass, both subject to radiation pressure.
QRPN can be separated in two contributions: static force and corresponding shift of the test mass's position (if the light powers are unequal on two sides of the test mass), and a noisy force, coming from quantum fluctuations of the light field.
The total force can be defined as a sum of all contributions:
$$\mathbb{F}_{\mathrm{rp}} = \sum \pm\mathcal{I}_i/c,$$
where $\mathcal{I}_i$ is the intensity of $i$-th field falling on the oscillator, and the sign of intensity is defined by the direction of travel.

There are two equivalent ways of calculating the QRPN: using two-photon and sideband picture.

\noindent \textit{In the sideband picture}:
\begin{multline}
\mathbb{F}_{\mathrm{rp}}/\hbar k_p = \mathbf{E}^\dagger\sigma_3\mathbf{e}(\omega_p + \Omega) + \mathbf{F}^\dagger\sigma_3\mathbf{f}(\omega_p + \Omega) + \\ + \mathbf{E}^T\sigma_3\mathbf{e}^\dagger(\omega_p - \Omega) + \mathbf{F}^T\sigma_3\mathbf{f}^\dagger(\omega_p - \Omega)
\end{multline}
\textit{In the two-photon picture:}
\begin{equation}
\mathbb{F}_{\mathrm{rp}}/\hbar k_p = \mathcal{E}_1^T \mathcal{e}_1(\Omega) - \mathcal{E}_2^T \mathcal{e}_2(\Omega) + \mathcal{F}_1^T \mathcal{f}_1(\Omega) - \mathcal{F}_2^T \mathcal{f}_2(\Omega)
\end{equation}
Here, we use the two-photon picture, since it makes it easier to compute the spectral densities.
We first compute the average intra-cavity fields and simplify the denominator in a single-mode approximation, but keep the numerator unexpanded (this will be done at a later stage):
\begin{align}
&E_1 = -A_1 \frac{T e^{i k_p L/2} \left(1 -i T e^{ik_p L}\right)}{\mathcal{D}(0)} = A_1\frac{Te^{i \theta/2}\left(1-i t R e^{-i\theta}\right)}{2 r \tau \gamma}\\
&E_2 = -A_1 \frac{r R T e^{3ik_pL/2}}{\mathcal{D}(0)}=A_1 \frac{r R T e^{-i\theta/2}}{2r\tau \gamma}\\
&F_1 = r E_1 + it E_2 = A_1 \frac{r T e^{i\theta/2}}{2 r \theta \gamma}\\
&F_2 = r E_2 + it E_1 = A_1 \frac{T\left( it e^{i\theta/2} + R e^{-i\theta/2}\right)}{2r\tau\gamma}
\end{align}
For the quantum amplitudes, the equations in the sideband picture are:
\begin{align}
& e_1 = \frac{Te^{ikL/2}}{\mathcal{D}(\Omega)} \left( -a_1 (1-itRe^{ik L}) - a_2 r R e^{ikL} + 2ik_p rR \mathcal{G}_2(\Omega)x(\Omega) e^{ikL/2}\right)\\
& e_2 = \frac{Te^{ikL/2}}{\mathcal{D}(\Omega)} \left( -a_2 (1-itRe^{ik L}) - a_1 r R e^{ikL}+ 2ik_p rR \mathcal{G}_1(\Omega)x(\Omega) e^{ikL/2}\right)\\
& \mathcal{G}_1(\Omega) = \frac{1}{T}\left((1-itRe^{ikL})E_1 - rRE_2e^{ikL}\right)\\
& \mathcal{G}_2(\Omega) = \frac{1}{T}\left(-(1-itRe^{ikL})E_2 + rRE_1e^{ikL}\right)\\
& f_1 = r e_1 + it e_2 + 2 i k_p x r E_1 \\
& f_2 = r e_2 + it e_1 - 2 i k_p x r E_2
\end{align}

The amplitudes of the fields depend on the position $x(\Omega)$, and therefore the QRPN has a dynamical (position-dependent) contribution.
This contribution is called dynamical back-action or optical rigidity\,\cite{Danilishin2012}.
First, we find the position-independent part of the QRPN, setting $x(\Omega)=0$, and moving to two-photon quadratures:
\begin{align}\label{eq:ring:internal}
  &\mathcal{e}_1 = \frac{T e^{i\Omega\tau/2}}{2r\tau (\gamma-i\Omega)}\left[\mathcal{P}_1(\Omega)\mathcal{a}_1  + \mathcal{P}_2(\Omega) \mathcal{a}_2\right],\\
  &\mathcal{e}_2 = \frac{T e^{i\Omega\tau/2}}{2r\tau (\gamma-i\Omega)}\left[\mathcal{P}_1(\Omega)\mathcal{a}_2  + \mathcal{P}_2(\Omega) \mathcal{a}_1\right],\\
  &\mathcal{f}_1 = r \mathbb{I} \mathcal{e}_1 - it\sigma_2 \mathcal{e}_2,\\
  &\mathcal{f}_2 = r \mathbb{I} \mathcal{e}_2 - it\sigma_2 \mathcal{e}_1,\\
  &\mathcal{E}_1 = \frac{T}{2r\tau \gamma}\mathcal{P}_1(0)\mathcal{A}_1, \\
  &\mathcal{E}_2 = \frac{T}{2r\tau \gamma}\mathcal{P}_2(0)\mathcal{A}_1, \\
  &\mathcal{F}_1 = r \mathbb{I} \mathcal{E}_1 - it\sigma_2 \mathcal{E}_2,\\
  &\mathcal{F}_2 = r \mathbb{I} \mathcal{E}_2 - it\sigma_2 \mathcal{E}_1,
\end{align}
where we defined the frequency-dependent matrices:
\begin{align}
&\mathcal{P}_1 =
\begin{bmatrix}
\cos \frac{\theta}{2} - t R e^{i\Omega\tau} \sin \frac{\theta}{2} &  - \sin\frac{\theta}{2} + t R e^{i\Omega\tau} \cos \frac{\theta}{2}\\
\sin \frac{\theta}{2} - t R  e^{i\Omega\tau}\cos \frac{\theta}{2} & cos \frac{\theta}{2} - t R e^{i\Omega\tau} \sin \frac{\theta}{2}
\end{bmatrix},\\
&\mathcal{P}_2 =  r R e^{i\Omega\tau}
\begin{bmatrix}
\cos \frac{\theta}{2} & \sin \frac{\theta}{2}\\
-\sin \frac{\theta}{2} & \cos \frac{\theta}{2}
\end{bmatrix}.
\end{align}
Taking these into account, we compute the contributions to the QRPN from the reflected fields:
\begin{align}
\mathcal{F}^T_1 \mathcal{f}_1 &= \left( r \mathbb{I} \mathcal{E}_1 - it\sigma_2 \mathcal{E}_2 \right)^T \left(r \mathbb{I} \mathcal{e}_1 - it\sigma_2 \mathcal{e}_2\right) =\nonumber\\
&= r^2 \mathcal{E}_1^T\mathcal{e}_1 + t^2 \mathcal{E}_2^T\mathcal{e}_2 + i r t \left( \mathcal{E}_1^T\sigma_2\mathcal{e}_2 - \mathcal{E}_2^T\sigma_2\mathcal{e}_1\right),\\
\mathcal{F}^T_2 \mathcal{f}_2 &= r^2 \mathcal{E}_2^T\mathcal{e}_2 + t^2 \mathcal{E}_1^T\mathcal{e}_1 + i r t \left( \mathcal{E}_2^T\sigma_2\mathcal{e}_1 - \mathcal{E}_1^T\sigma_2\mathcal{e}_2\right),
\end{align}
which leads to the simplified expression for the QRPN:
\begin{equation}
\mathbb{F}_{\mathrm{rp}} = 2 \hbar k_p \left(r^2(\mathcal{E}_1^T\mathcal{e}_1 - \mathcal{E}_2^T\mathcal{e}_2) + i r t(\mathcal{E}_1^T\sigma_2\mathcal{e}_2 - \mathcal{E}_2^T\sigma_2\mathcal{e}_1)  \right).
\end{equation}
Once again choosing the phase of the input field such, that $\mathcal{A}_1 = \{A, 0\}^T/\sqrt{2}$,
we arrive at the final expression for the QRPN in the single mode approximation:
\begin{multline}\label{eq:ring:frp}
  \mathbb{F}_{\mathrm{rp}}(\Omega) = \frac{\hbar k_p A}{\sqrt{2} (\gamma-i\Omega)} \left[ \left(r^2 - t^2\right) \left(2 \gamma - i\Omega\right)a^c_1(\Omega) + i rt\Omega a^s_1(\Omega) - \right. \\ - \left. i\left(r^2 - t^2\right) \Omega a^c_2(\Omega)  + i r t\left(2 \gamma - i\Omega\right) a^s_2(\Omega)\right].
\end{multline}
This allows us to compute the spectral density of the QRPN for the case of a perfectly reflective test mass:
\begin{equation}
S_{\mathrm{rp}}(\Omega) = 2\gamma^2\frac{\hbar \omega_p I_{\mathrm{in}}}{c^2 (\gamma^2 + \Omega^2)} +  \Omega^2\frac{\hbar \omega_p I_{\mathrm{in}}}{c^2 (\gamma^2 + \Omega^2)} = S_{\mathrm{rp}}^x(\Omega) +  S_{\mathrm{rp}}^v(\Omega).
\end{equation}
This spectral density has two contributions: position $S_{\mathrm{rp}}^x(\Omega)$ and speed $S_{\mathrm{rp}}^v(\Omega)$.
This corresponds to the two detection channels.
Since every measurement produces a back action, the position port must have its contribution, as well as the speed one.

While the position contribution to radiation-pressure force limits the speedmeter enhancement, it can be minimized by one of the several ways.
Firstly, it can be subtracted via optimal combination of the two output fields, as we show in the next section. 
Second, the two input light fields can be entangled such, that $(r^2-t^2)a_1^c = -irt a_2^s$, in which case the position-dependent part in Eq.\,\ref{eq:ring:frp} gets canceled.
Implementing such correlations could be done through two-mode squeezing, but can be challenging in practice.
Secondly, if $r=t$, it is sufficient to inject squeezed light to suppress $a_2^s$ to achieve speedmeter sensitivity.
These approaches require further study into their detailed implementation and feasibility. 

Another important fact is that the QRPN in the HRC is significantly smaller than the noise in a conventional Fabry-Perot cavity:
\begin{equation}
S_{\mathrm{rp, FP}}(\Omega) = \frac{4\hbar I_{\mathrm{in}} \omega_p}{L^2 (\gamma^2 + \Omega^2)}.
\end{equation}
This is expected: since the shot noise is increased compared to the Fabry-Perot cavity, the radiation-pressure must be reduced to obey the Heisenberg uncertainty relation (for a minimum uncertainty state).
Fundamentally, the origin of this reduction is exactly the symmetry of the setup that allows the speed measurement: most of the radiation pressure is canceled coherently on the round trip.
Only a small fraction of the signal that leaks through the front mirror, is not canceled.

Intra-cavity light field also depends on the position of the test mass, which causes a position-dependent part of the QRPN.
From Eqs.\,\ref{eq:ring:internal}, we obtain this position-dependent part in the single-mode approximation:
\begin{equation}
\mathbb{F}_{\mathrm{rp}}^{x} = \frac{4 \omega_p I_{in} t}{c L \gamma } x = \mathcal{K}_{\mathrm{os}} x,
\end{equation}
where $\mathcal{K}_{\mathrm{os}}$ is the optical spring constant.
The dynamics of the mirror is described by:
\begin{equation}
  M\ddot{x}(t) + 2\gamma_m \dot{x}(t) + M\omega_m^2 x(t) = F_{\mathrm{rp}}(t) + F_{\mathrm{rp}}^{x}(t) + F_{\mathrm{cl}}(t)
\end{equation}
where $\omega_m$ is the mechanical frequency, $\gamma_m$ is the mechanical linewidth, $M$ is test mass's mass, $F_{\mathrm{cl}}$ is the random force caused by any classical forces, such as thermal fluctuations in the test mass.
Taking into account the change in dynamics due to the QRPN, the dynamics changes:
\begin{equation}
  M\ddot{x}(t) + 2\gamma_m \dot{x}(t)  + \left(M\omega_m^2 - \mathcal{K}_{\mathrm{os}}\right) x(t) = F_{\mathrm{rp}}(t) + F_{\mathrm{cl}}(t).
\end{equation}
Dynamical QRPN effectively introduces a shift to the mechanical frequency, thus its name -- "optical spring".
This optical spring is different from the optical spring arising in a Fabry-Perot cavity, where additionally to changing the frequency, the QRPN introduces additional damping or anti-damping.
In a Fabry-Perot cavity optical spring arises only when the laser is detuned off the cavity's resonance frequency.
In the HRC the optical spring arises on resonance, and it only shifts the mechanical resonance frequency, without introducing any damping of anti-damping.
However, this optical spring is rather weak: for the experimental parameters of a HRC presented in the experimental section of this chapter, the shift due to the optical spring is on the order of  10\,kHz to the mechanical frequency of 395.2\,kHz.
We explored the properties of the optical spring in detail in Ref.\,\cite{liCoherentCouplingCompleting2019}.

\subsection{Optimal readout}

\begin{figure}
  \includegraphics[width=1\textwidth]{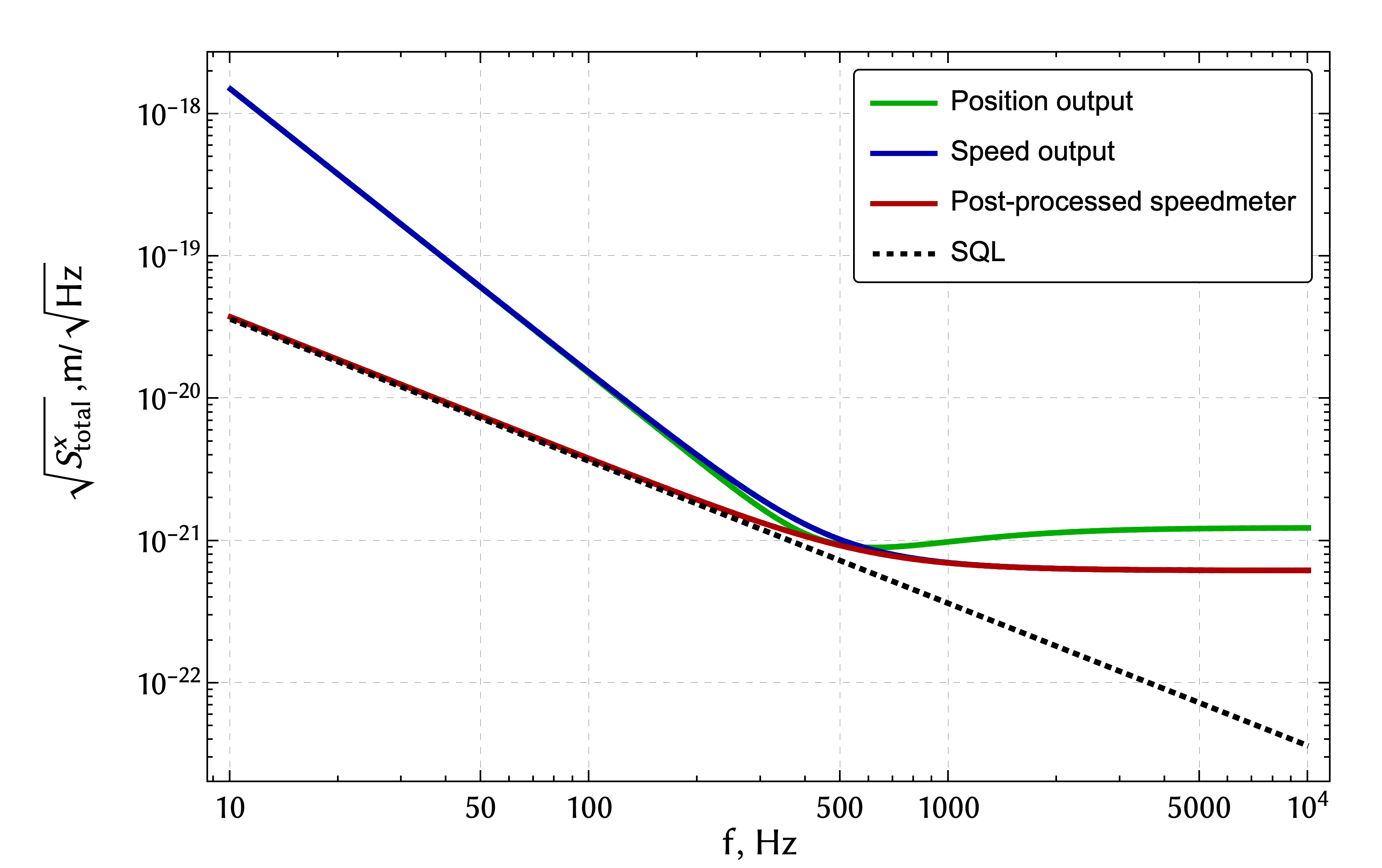}
  \caption{Quantum-limited displacement sensitivity of the km-scale HRC on the position and speed output ports, as well as the post-processed speedmeter sensitivity.
Without post-processing, the sensitivity is dominated by the position-dependent part of QRPN.
Measuring the position output and optimally processing the combination of the two signals allows to achieve speedmeter scaling.}
  \label{fig:conditional}
\end{figure}

In this section, we discuss the possibility to combine two output readout signals to achieve speedmeter behavior for QRPN.
From Eq.\,\ref{eq:ring:frp} we can write for the fully reflective test mass:
\begin{equation}
    \mathbb{F}_{\mathrm{rp}}(\Omega) = \frac{\hbar k_p A}{\sqrt{2} (\gamma-i\Omega)} \left[ \left(2 \gamma - i\Omega\right)a^c_1(\Omega) - i\Omega a^c_2(\Omega)\right].
\end{equation}
Here we see that the amplitude quadrature from both input ports contributes to the radiation-pressure force. 
We need to cancel the position-dependent (i.e. proportional to $\gamma$) part of it, which can be done by measuring the output amplitude quadrature at the position port, which contains no information about the test mass displacement:
\begin{equation}
  b_1^c = \frac{\gamma}{\gamma-i\Omega}a_1^c - \frac{i\Omega}{\gamma - i\Omega}a_2^c.
\end{equation}
Then the optimal combination of the two measurement records would allow to suppress the contribution of the position-dependent radiation-pressure noise:
\begin{equation}
  b_2^{\rm opt} = b_1^s + g(\Omega)b_2^c,
\end{equation}
where $g(\Omega)$ is a frequency-dependent filtering function. 
The total output then becomes, including the QRPN-induced displacement:
\begin{multline}
  b_2^{\rm opt} = \frac{\gamma}{\gamma-i\Omega}a_2^s - \frac{i\Omega}{\gamma - i\Omega}a_1^s - \\ i\Omega \frac{k_p A_1}{\gamma - i\Omega} \frac{\hbar k_p A_1}{\sqrt{2} M\Omega^2(\gamma-i\Omega)} \left[ \left(2 \gamma - i\Omega\right)a^c_1(\Omega) - i\Omega a^c_2(\Omega)\right] + \frac{g(\Omega)}{\gamma-i\Omega}\left[\gamma a_1^c - i\Omega a_2^c\right] = \\ \frac{\gamma}{\gamma-i\Omega}a_2^s - \frac{i\Omega}{\gamma - i\Omega}a_1^s + \frac{a_1^c}{\gamma-i\Omega}\left[K(\Omega)\left(2\gamma - i\Omega\right) + g(\Omega)\gamma\right] -i\Omega \frac{a_2^c}{\gamma - i\Omega}\left(K(\Omega) + g(\Omega)\right),
\end{multline} 
where we defined the coefficient 
$$K(\Omega) = i\Omega \frac{\hbar k_p^2 A_1^2}{\sqrt{2}M\Omega^2(\gamma-i\Omega)}.$$
An obvious choice of the filtering function $g = -2K$ removes the position-dependent part from the readout:
\begin{equation}
  b_2^{\rm opt} = \frac{\gamma}{\gamma-i\Omega}a_2^s - \frac{i\Omega}{\gamma - i\Omega}a_1^s - i\Omega \frac{K(\Omega) (a_1^c - a_2^c)}{\gamma-i\Omega},
\end{equation}
and the resulting spectral density contains no position contribution to the QRPN, as shown in Fig.\,\ref{fig:conditional}. 
In practice, this requires exact knowledge of the parameters of the system to be able to achieve the optimal filtering.
The sensitivity could surpass the SQL if an optimal homodyne angle is chosen for the speed output.

\subsection{Towards the design of a large-scale detector}

\begin{figure}
  \includegraphics[width=0.7\textwidth]{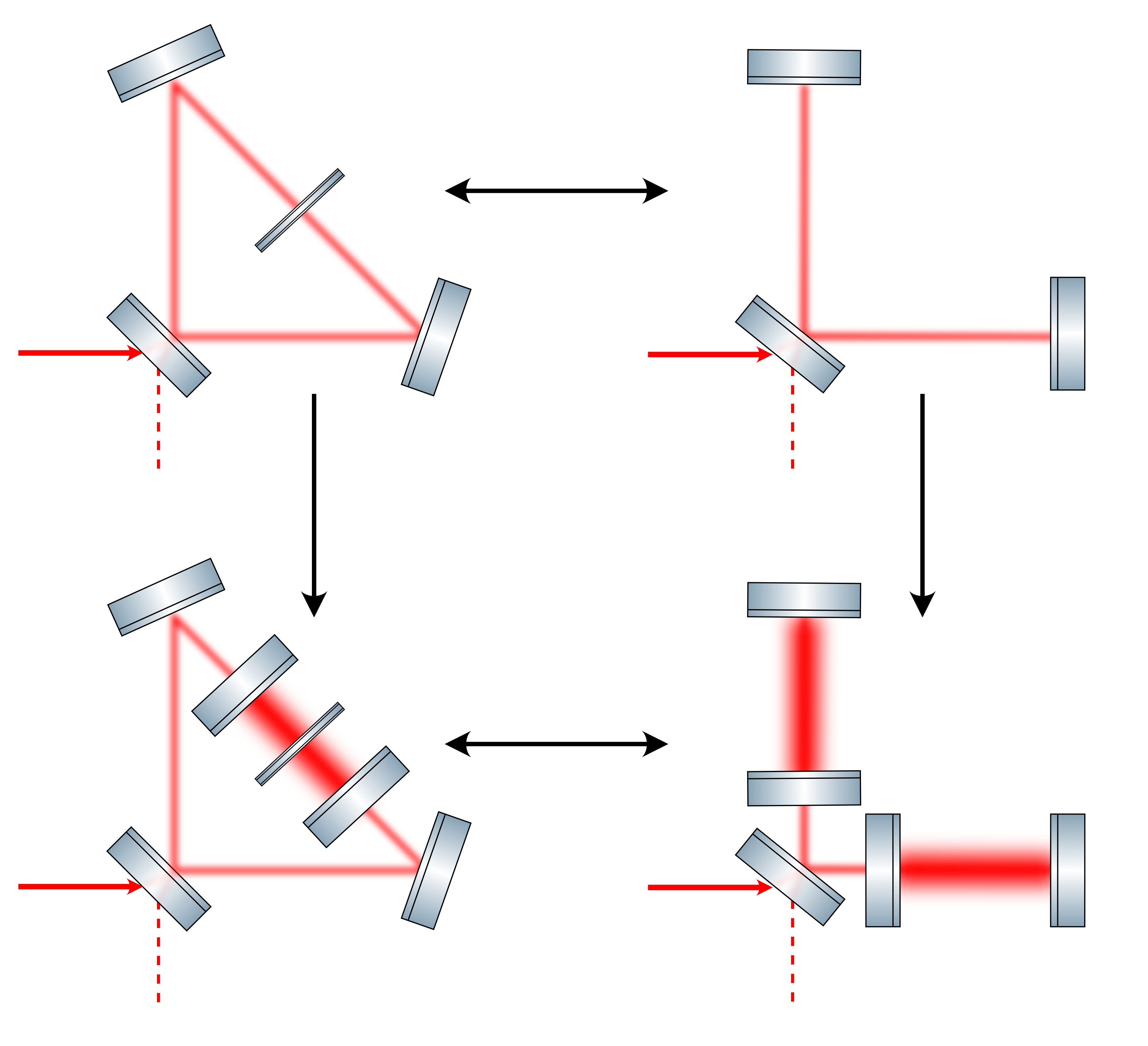}
  \caption[A path towards ring-cavity-based GW detector.]{A path towards ring-cavity-based GW detector.
  The HRC with a perfectly reflective test mass (top left) is invariant to the L-shaped cavity (top right).
  Placing a Fabry-Perot cavity around the test mass (bottom left) enhances the light power as well as the signal.
  So do the arm cavities in the L-shaped topology (bottom right).
  The resulting sensitivity is compatible with traditional speedmeter topologies (e.g.\,a Sagnac speedmeter).}
  \label{fig:7}
\end{figure}

The HRC topology offers interesting properties, that might be used for future gravitational wave detectors.
In particular, it would be possible to implement the ring-cavity type design already in the current generation of GWOs.
For that purpose the central beamsplitter has to be "rotated" by 90 degrees to form a front mirror that couples the two arms of the interferometer, see Fig\,\ref{fig:7}.
Unlike a simple HRC, in this configuration the signal will be amplified by the arm cavities.
Such a detector should feature the full enhancement to the sensitivity from the arm cavities, and the speedmeter signal on the output of the detector. 
We do not provide the full derivation of the sensitivity here, leaving this as a subject of future work.

However, there are challenges associated with such design.
Firstly, to achieve the speedmeter scaling of the radiation-pressure noise, an optimal filtering of the signals from both output ports is required, as we discuss in the previous section and the Appendix.
Secondly, since the two end test masses move independently, the dark (speed) port contains not only a speed signal from the differential motion of the mirrors, but also common motion noise, which contaminates the sensitivity, since it contains no GW signal.
Usually in a standard Michelson interferometer, this common mode is canceled naturally.
In the L-shaped detector, this is not the case, and special measures would have to be considered for avoiding the influence of this common mode contamination.
However, the analysis performed in Ref.\,\cite{guoMergingLshapedResonator2024, guoSensingControlScheme2023, zhangGravitationalWaveDetectorPostmerger2023a}, where a similar setup was developed with a different motivation, shows that it is still possible to achieve required stability of the setup.

Overall, while the presented configuration is non-standard in terms of its features, it offers potential benefits from using only one field in one polarization in a standard topology, which differs it from the majority of speedmeter proposals. 
Further research is needed to assess other feasibility aspects of this approach.

\subsection{Membrane speedmeter response}
\begin{figure}
  \centering
  \includegraphics[width=0.8\textwidth]{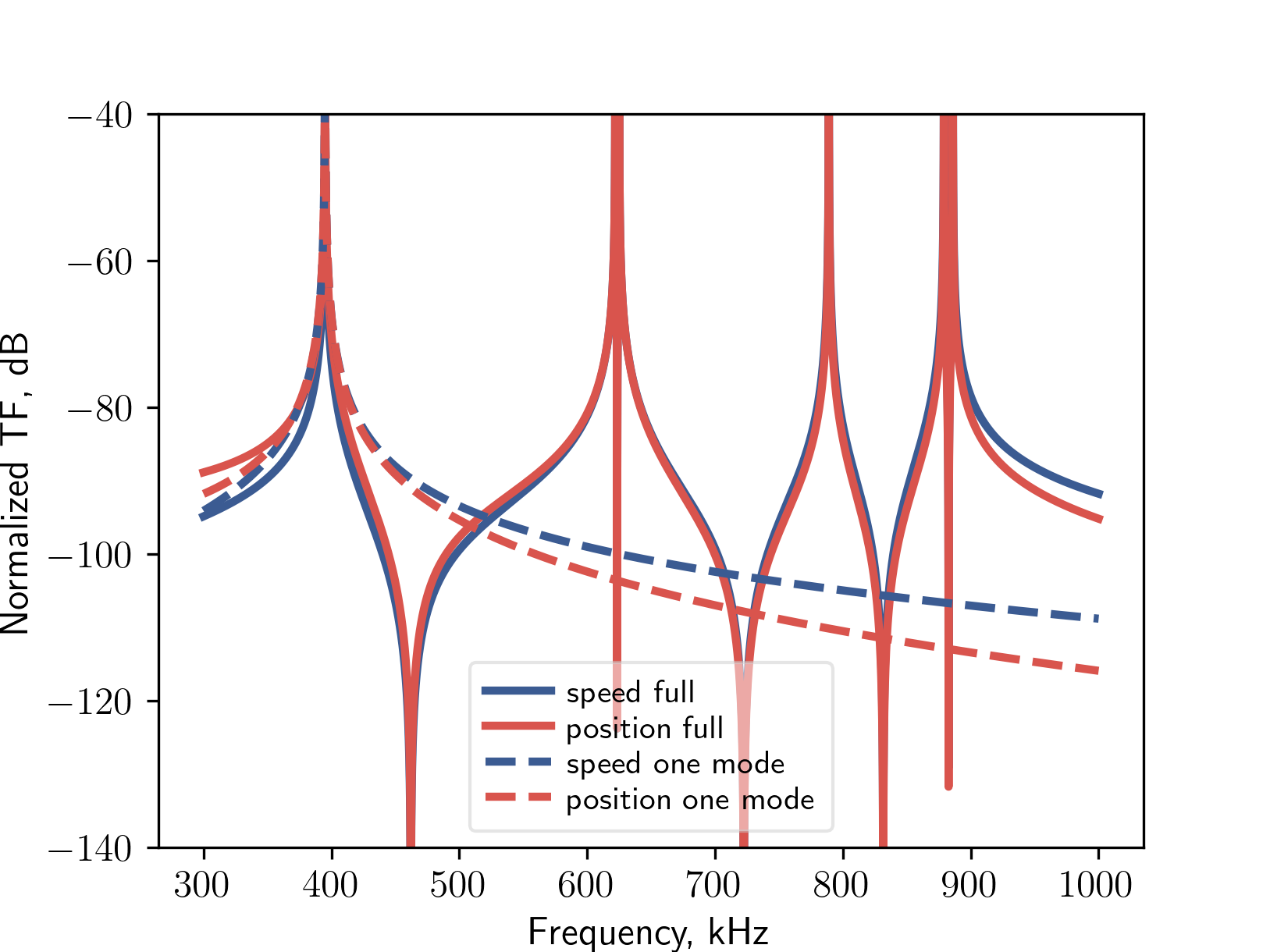}
  \caption[Theoretical calculation of the signal transfer function for an external force acting on the HRC's membrane as measured by the BHD.]
  {
  Theoretical calculation of the signal transfer function for an external force acting on the HRC's membrane as measured by the BHD. The transfer functions are normalized to their corresponding peak values. Shown are transfer function for one mechanical mode, which shows characteristic response at high frequency (dashed) and a multi-mode case of a real membrane. The free-mass regime above the resonance of the fundamental mode is spoiled by the interaction with the next high-order mode, which makes the observation of difference between the speed and position response unfeasible in that frequency regime. The effective mass of each mode is not taken into account here.
  }
  \label{fig:A8}
\end{figure}

As discussed in the main text, the mechanical membrane turned out to be not a suitable test mass for observing all speedmeter signatures. 
The reason for that are the high-order mechanical modes that contribute destructively to the signal.
The frequencies of these modes are defined by the geometry of the membrane:
\begin{equation}
  f_{m,n} = \sqrt{\frac{T}{4\rho}}\sqrt{\frac{m^2}{X^2}+\frac{n^2}{Y^2}},
\end{equation}
where $T=800\times10^6$\,Pa is the stress of the membrane, $\rho\sim 2.7$\,\si{kg/m^3} is the mass density, (m, n) are the index numbers of the modes, and $(X,Y)$ are the dimensions of the membrane.
At frequencies between two neighboring modes, transfer function experiences a minimum where the response of two modes destructively interferes. 
This spoils the expected speedmeter behavior which is expected to be observed exactly at those frequencies.
Fig.\,\ref{fig:A8} shows the calculation of the transfer function based on Eq.\,\ref{eq:ring:transfer}, modeling the mechanical response of each mode as a standard mechanical oscillator.

\section{Appendix: setup characterization}

\subsection{Optical cavity}
\begin{figure}
  \includegraphics[width=1\textwidth]{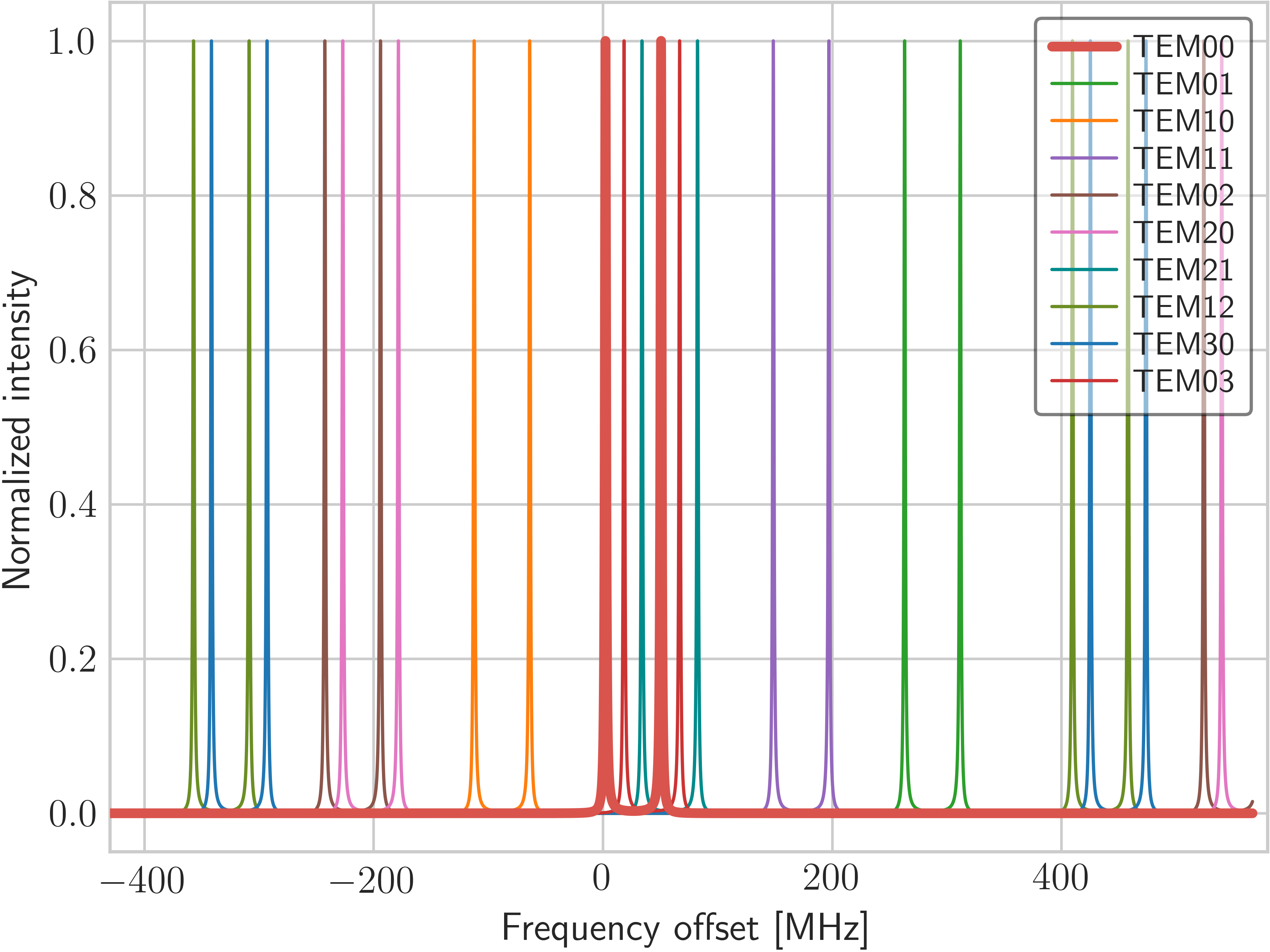}
  \caption[Simulation of resonances of the higher order TEM modes in the HRC.]{Simulation of resonances of the higher order TEM modes in the HRC.
  The design length was selected such that there was a minimal overlap between the TEM00 modes and the higher order modes, such that only one became resonant at a selected frequency.
  All modes are normalized to their peak intensity. Image was produced using FINESSE software package.}
  \label{fig:ring:TEM}
\end{figure}

\begin{figure}
  \centering
  \includegraphics[width=1\textwidth]{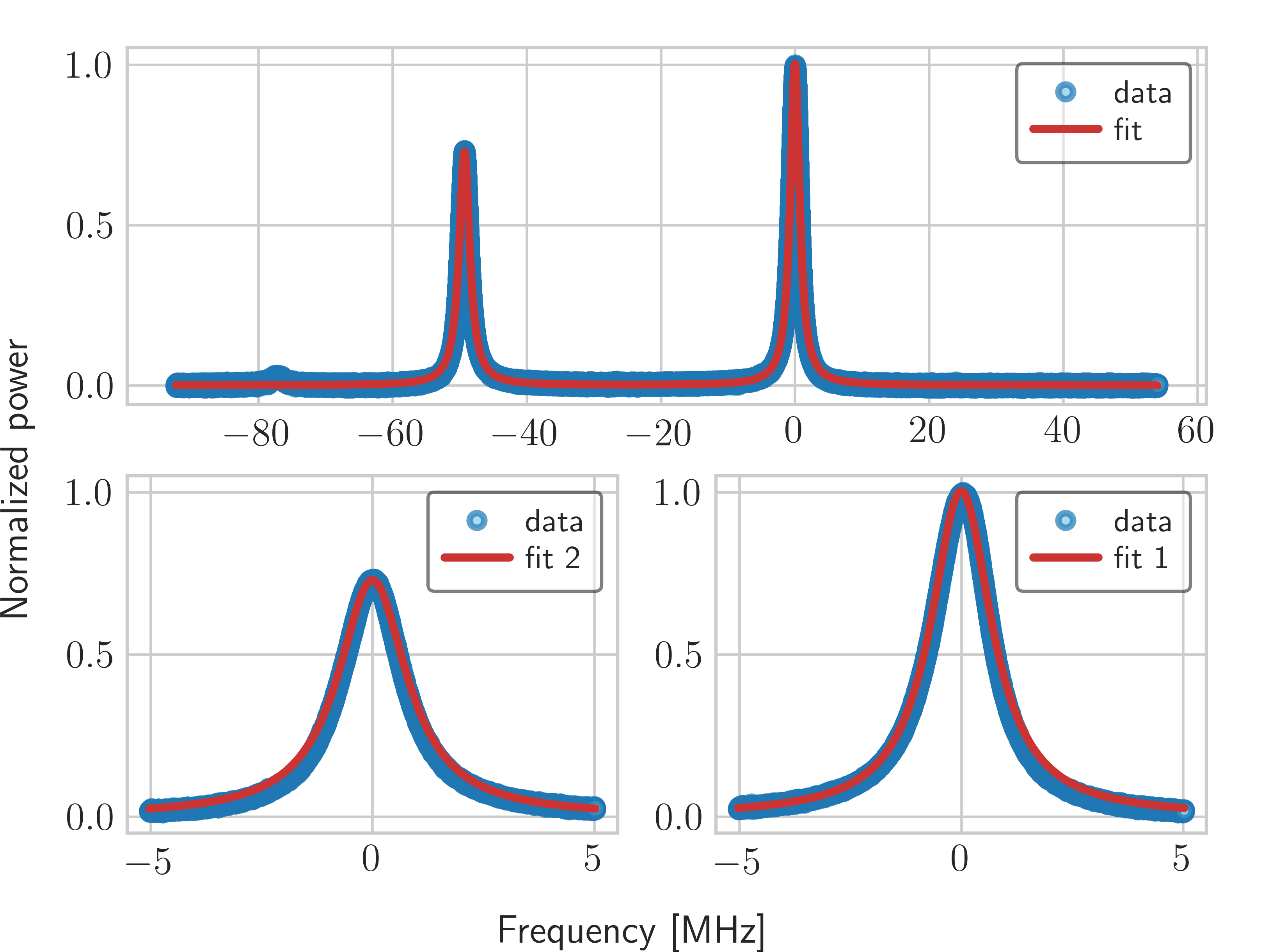}
  \caption{Measurement of the linewidth and the mode splitting of the HRC with a membrane.
  The frequency of the laser was scanned continuously, and the time trace of the power changes in the position port was recorded.
  The frequency scale was calibrated using a PDH signal.
  The power was normalized to the peak power of the larger peak.
  The parameters acquired from the fit: $\gamma_1 = 0.84\pm0.01$\,MHz, $\gamma_2 = 0.95\pm0.01$\,MHz, $\Delta \omega_c = 49.28$\,MHz, $r=3.9\%$.
  The experimentally measured linewidth matched the theoretical prediction of $\gamma = 0.84$\,MHz, based on the mirrors' parameters.
  The power of one of the peaks differed from the other one due to a contribution of a higher-order optical mode that became degenerate with the TEM00 mode, so the corresponding peak was higher, but also broader.}\label{fig:ring:linewidth}
\end{figure}

The cavity length was optimized to minimize the effect of high-order optical modes on the main resonance, see Fig.~\ref{fig:ring:TEM}.
The cavity parameters were experimentally verified by measuring the linewidth of the cavity (with and without the membrane), see Fig.~\ref{fig:ring:linewidth}.
The optical design parameters are summarized in Table~\ref{tab:ring:exp_pars}.

\begin{table}[t]
  \centering
  \begin{tabular}{|l|l|}
    \hline
    \hline
    Cavity parameters & \\
    \hline
      Length & 39.1\,cm\\
      Front mirror power transmission &  0.01\\
      Back mirror power transmission & 0.0007\\
      Back mirror radius of curvature & 25\,cm\\
      Back mirror angle of incidence & $10.5^\circ$\\
      Cavity waist $w_x$ & 221$\mu$m\\
      Cavity waist $w_y$ & 229$\mu$m\\
      Cavity linewidth peak 1 & $0.84\pm0.01$\,MHz\\
      Cavity linewidth peak 2 & $0.95\pm0.01$\,MHz\\
      Cavity resonance splitting & $\sim$53.5\,MHz\\
      Light power input & 0.01\,mW\\
      Light power detection & \SI{100}{\uW}\\
    \hline
    Membrane parameters & \\
    \hline
      Membrane reflectivity & 4.6\%\\
      Membrane size & 1\,mm $\times$ 1\,mm\\
      Membrane clipping & $ < 0.002$\%\\
      Membrane frequency & 395.2\,kHz\\
      Membrane Q factor & $4.6\times10^5$\\
      \hline
      \hline
  \end{tabular}
  \vskip3pt
  \caption[Main experimental parameters of our HRC system.]{Main experimental parameters of our HRC system.}\label{tab:ring:exp_pars}
\end{table}

The laser was stabilized to the cavity resonance by the Pound-Drever-Hall technique\,\cite{Drever1983,Black2001}.
For that purpose a phase modulation sideband was reflected off the cavity, and the beat between the main field and the sideband was detected on a separate photodiode.
After demodulation at the sideband frequency, the resulting error signal was fed back to the servo controller with appropriately designed integrators and filters, which produced a control signal for the piezoactuator acting on the length of a diode cavity in the laser.

\subsection{Membrane motion}
The membrane was positioned inside the HRC close to the front mirror (since the macroscopic position of the membrane did not have an effect on the resonance structure).
The whole cavity was placed inside a vacuum chamber with residual pressure of $10^{-7}$\,mbar.
The vacuum could be maintained with only the high-speed ion-getter pump, which minimized the coupling of acoustic noise to the cavity.
The cavity was isolated from the vibrations of the table by two-stage passive isolation made of alternating layers of steel sheets and Viton feet.
The optical table itself was floating on air-filled vibration isolation feet.

When the cavity was brought to resonance, most of the light power was reflected directly back to the position port.
Only a small fraction of power went into the speed port (due to the non-perfectly reflective back mirror).
The signal in the position port was then separated from the incoming field with the help of a Faraday rotator and a PBS.

The signal from the resonating HRC was sent to a balanced homodyne detector, from two ports: position and speed.
These two signals were overlapped on a PBS, allowing to switch between detecting one or another without changing the alignment by blocking one path and adjusting the polarization with a half-wave plate.
This PBS also allowed to adjust the light power to match the signal strength in both ports (since position port has most of the input power, it has to be reduced significantly).
The homodyne was adjusted to achieve good visibility and mode overlap (although it was not critical for the experiment), with the use of an additional diagnostic cavity.
The phase of the local oscillator was actively stabilized to the phase quadrature, where the membrane signal reaches maximal strength.

\begin{figure}
  \centering
  \includegraphics[width=1\textwidth]{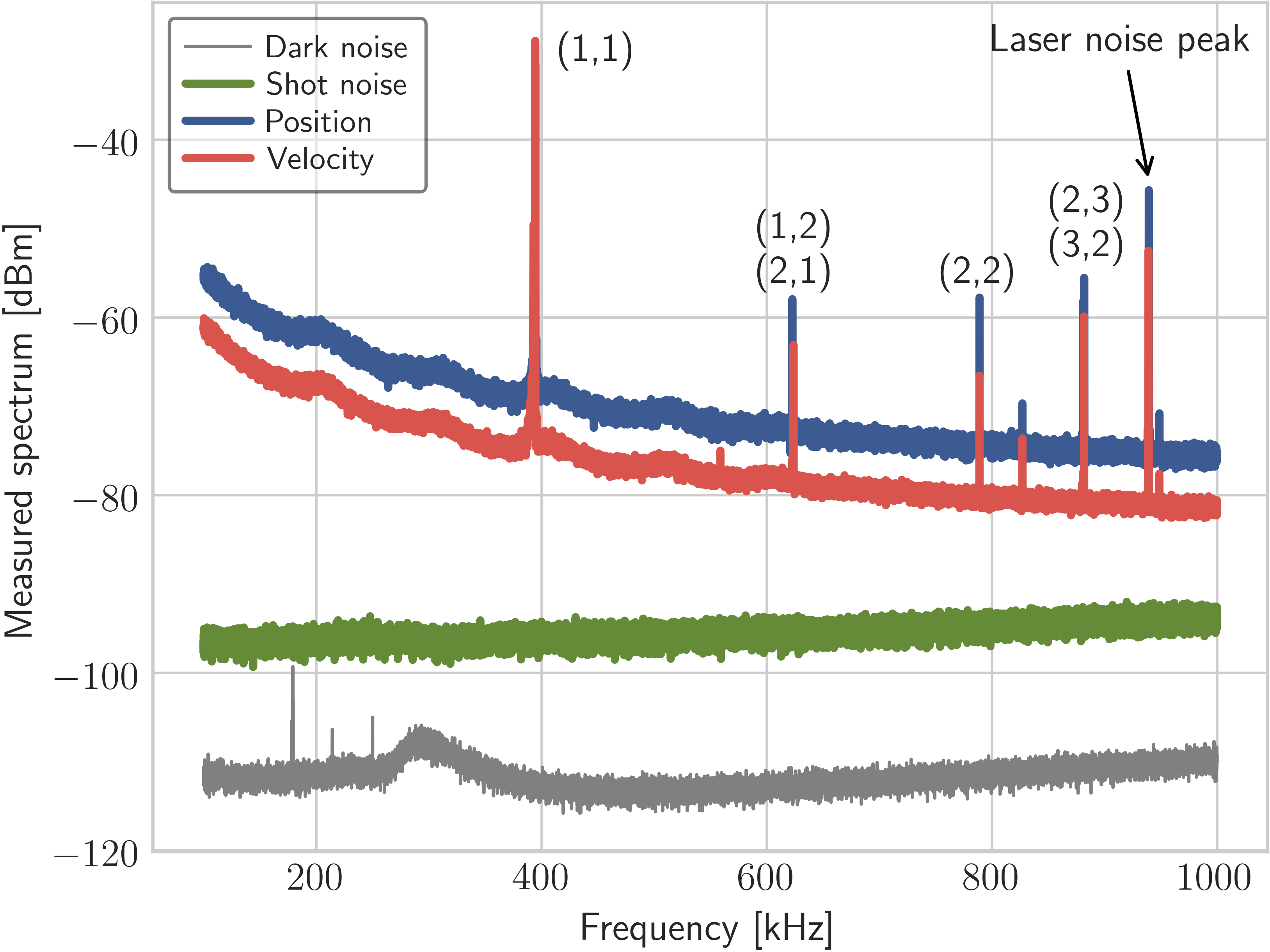}
  \caption[Spectral density of noise and membrane motion.]{Spectral density of noise and membrane motion.
  The membrane peaks were produced by thermal motion of the membrane.
  The fundamental (1,1) mode is at $395.2$\,kHz.
  The higher order modes at higher frequencies fit the expected frequencies.
  The measurement was limited by the phase noise of the laser, which was significantly higher than the shot noise.}
  \label{fig:ring:peaks}
\end{figure}

A typical measured spectrum is presented in Fig.\ref{fig:ring:peaks}.
This measurement was largely limited by the laser phase noise.
The membrane motion was produced by thermal Brownian noise exciting the resonance frequencies of the membrane.
In the experimental data, the fundamental mode $f_{11}$ at 395.2\,kHz is well visible, together with the higher-order modes at (622, 625, 789, 879, 886)\,kHz.

\begin{figure}
  \centering
  \includegraphics[width=1\textwidth]{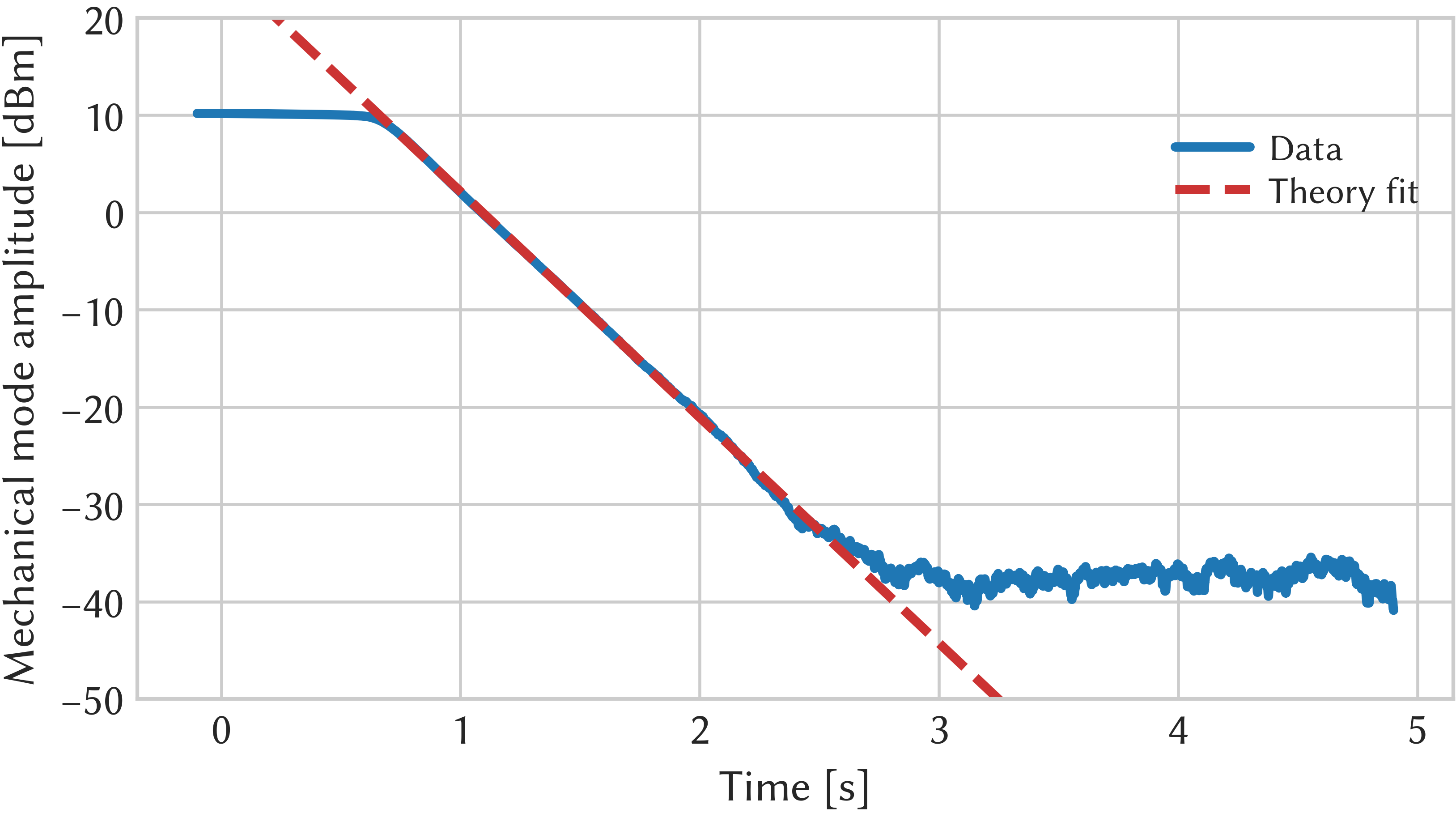}
  \caption[Ringdown measurement of the quality factor of the membrane's fundamental mode.]{Ringdown measurement of the quality factor of the membrane's fundamental mode.
  The membrane motion was measured without a cavity, i.e.\,the front mirror of the cavity was replaced by a blank substrate, and the reflection off the membrane was sent directly to the homodyne detector.
  The membrane was then excited at the resonance frequency of 395.2\,kHz, and the time trace of its motion was recorded.
  When the excitation was removed, the motion ringed down at the membrane's damping rate, which was fitted and a quality factor of $4.6 \times 10^5$ was computed after multiple repetitions of the measurement.}
  \label{fig:ring:k395}
\end{figure}

In order to characterize the membrane, we performed a series of ring-down measurements.
For this measurement, a small piezoactuator was attached to the membrane holder, and the membrane was excited at its resonance frequency by applying a sinusoidal voltage to the piezo.
By repeating this measurement 100 times, we acquired an averaged trace, presented in Fig.~\ref{fig:ring:k395}.
From this trace we computed the quality factor of the membrane by fitting an exponential ringdown to the data:
\begin{equation}
  X(t) = X(t_0)e^{- \omega_m (t-t_0)Q^{-1}} \quad \Rightarrow \quad Q = \frac{10\omega_m(t-t_0)}{\left(X_{\mathrm{dB}}(t_0)-X_{\mathrm{dB}}(t)\right)\log 10}
\end{equation}
The resulting quality factor of $Q\sim4.6\times10^5$ is typical for this type of membrane\,\cite{Zwickl2008}.

\end{document}